\documentclass[aps,prx,reprint,showpacs,superscriptaddress,notitlepage,floatfix]{revtex4-2}

\usepackage[pdftex]{graphicx}
\usepackage[pdftex]{epsfig}
\usepackage{epstopdf}
\usepackage{color}
\usepackage{amssymb,amsmath}
\usepackage{graphicx}
\usepackage{dcolumn}
\usepackage{multirow}
\usepackage{hyperref}
\usepackage{siunitx}
\usepackage[dvipsnames]{xcolor}
\usepackage{xspace}
\usepackage{booktabs}
\usepackage{txfonts}

\usepackage{tikz}
\usetikzlibrary{quantikz}

\usepackage{mathtools}

\newcommand{\qc}{\mathcal{Q}}

\begin{document}

\title{Benchmarking quantum computers with any quantum algorithm}
\author{Stefan K. Seritan}
\affiliation{Quantum Performance Laboratory, Sandia National Laboratories, Albuquerque, NM 87185, USA and Livermore, CA 94550, USA}
\affiliation{Quantum Applications and Algorithms Collaboratory, Sandia National Laboratories, Albuquerque, NM 87185, USA and Livermore, CA 94550, USA}
\author{Aditya Dhumuntarao}
\affiliation{Quantum Performance Laboratory, Sandia National Laboratories, Albuquerque, NM 87185, USA and Livermore, CA 94550, USA}
\author{Aidan Q. Wilber-Gauthier}
\affiliation{Quantum Performance Laboratory, Sandia National Laboratories, Albuquerque, NM 87185, USA and Livermore, CA 94550, USA}
\affiliation{Quantum Applications and Algorithms Collaboratory, Sandia National Laboratories, Albuquerque, NM 87185, USA and Livermore, CA 94550, USA}
\author{Kenneth M. Rudinger}
\affiliation{Quantum Performance Laboratory, Sandia National Laboratories, Albuquerque, NM 87185, USA and Livermore, CA 94550, USA}
\author{Antonio E. Russo}
\affiliation{Quantum Applications and Algorithms Collaboratory, Sandia National Laboratories, Albuquerque, NM 87185, USA and Livermore, CA 94550, USA}
\author{Robin Blume-Kohout}
\affiliation{Quantum Performance Laboratory, Sandia National Laboratories, Albuquerque, NM 87185, USA and Livermore, CA 94550, USA}
\author{Andrew D. Baczewski}
\affiliation{Quantum Applications and Algorithms Collaboratory, Sandia National Laboratories, Albuquerque, NM 87185, USA and Livermore, CA 94550, USA}
\author{Timothy Proctor}
\affiliation{Quantum Performance Laboratory, Sandia National Laboratories, Albuquerque, NM 87185, USA and Livermore, CA 94550, USA}

\begin{abstract}
Application-based benchmarks are increasingly used to quantify and compare quantum computers’ performance. However, because contemporary quantum computers cannot run utility-scale computations, these benchmarks currently test this hardware’s performance on “small” problem instances that are not necessarily representative of utility-scale problems. Furthermore, these benchmarks often employ methods that are unscalable, limiting their ability to track progress towards utility-scale applications. In this work, we present a method for creating scalable and efficient benchmarks from any quantum algorithm or application. Our \emph{subcircuit volumetric benchmarking} (SVB) method runs subcircuits of varied shape that are ``snipped out’’ from some target circuit, which could implement a utility-scale algorithm. SVB is scalable and it enables estimating a \emph{capability coefficient} that concisely summarizes progress towards implementing the target circuit. We demonstrate SVB with experiments on IBM Q systems using a Hamiltonian block-encoding subroutine from quantum chemistry algorithms.
\end{abstract}
 
\maketitle

\section{Introduction}\label{sec:introduction}
Advances in quantum computing hardware since 2015 have spurred increasing interest in benchmarking the performance of quantum computing systems \cite{Proctor2025-cd, Hashim2024-om, Lubinski2023-zy, Amico2023-ze, Proctor2021-wt,  Blume-Kohout2020-de, Erhard2019-wk, Cross2019-ku}. One approach is to run applications on the quantum computer, then measure and report the error in the results \cite{Proctor2025-cd, Hashim2024-om, Lubinski2023-zy}. This methodology underpins many application-based quantum computer benchmarking suites introduced in recent years \cite{Chen2022-dm, Tomesh2022-nu, Linke2017-mr, Wright2019-zj, Murali2019-my, Donkers2022-wt, Finzgar2022-aa, Mills2020-zh, Lubinski2023-zy, Lubinski2024-ci, Lubinski2023-mr, Chen2023-la, Benedetti2019-pp, Li2020-ry, Quetschlich2023-bg, Dong2021-gj, Martiel2021-vp, Van_der_Schoot2022-gv,Van_der_Schoot2023-vo, Cornelissen2021-yt, Georgopoulos2021-hh, Dong2022-ga, Chatterjee2025-lp}. But because contemporary quantum computers cannot yet solve practically useful problems, this approach has only enabled testing their ability to solve ``small'' problems using small quantum circuits (programs) that are far from the \emph{teraquop} (one trillion quantum operations, or \emph{quops}) regime expected to enable true quantum utility \cite{Proctor2025-cd,gidney2025factor,low2025fast,Rubin2024-gc}. It is unclear whether a quantum computer's performance on small problem instances, or on subroutines required to solve those small problems, can be used to assess technological progress towards the goal of quantum utility.

\begin{figure*}[ht!]
    \centering
    \includegraphics[width=\textwidth]{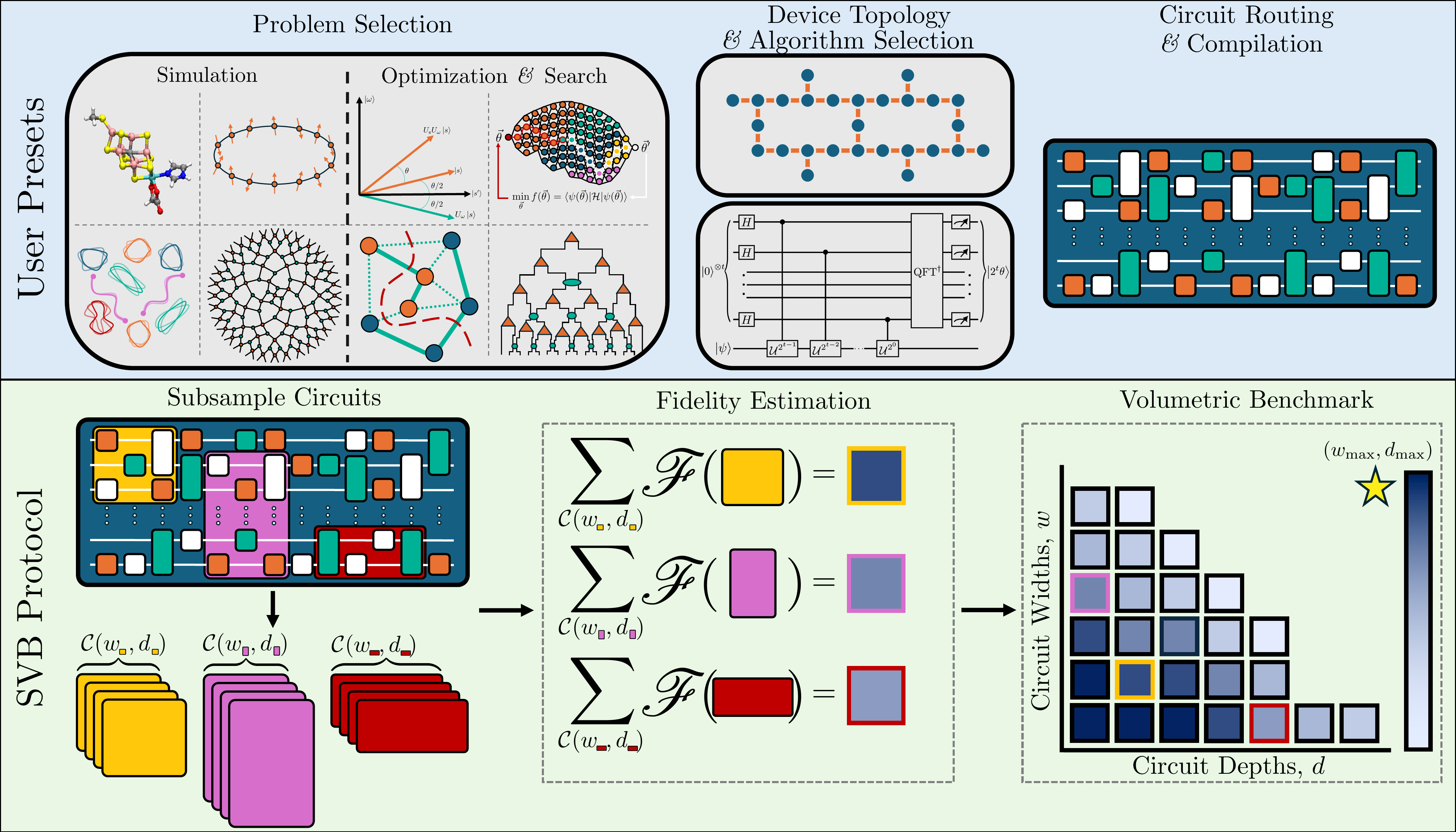}
    \caption{{\small\textbf{Subcircuit volumetric benchmarking}. A diagram of the subcircuit volumetric benchmarking (SVB) method that we introduce in this paper, which enables creating scalable and efficient benchmarks from quantum algorithms. The SVB method (green lower box) takes, as its input, a fully compiled \textit{target circuit} $c$. Although any target circuit can be used, we envisage $c$ implementing an interesting quantum algorithm, constructed by compiling an algorithm for an interesting computational problem (e.g., ``factor RSA-2048'' \cite{Proctor2025-cd}) so that it can be run on a particular quantum computer. It may be too large to execute on contemporary quantum computers, containing millions or trillions of gates.  SVB efficiently assesses how close a particular quantum computer is to being able to execute $c$ with a high probability of success. It does so by ``snipping out'' subcircuits of various shapes from $c$ (lower left panel), executing efficient experiments that enable estimating the process fidelity with which those subcircuits are executed (lower middle panel), and then summarizing performance in a volumetric plot (lower right panel) that shows how the subcircuits' average process fidelity decays with increasing circuit width (the number of qubits) and depth (the number of layers of gates). These volumetric plots provide a visual summary of how far a system is from successfully solving the problem of interest and can be used to quantify this ``distance'' in terms of a \emph{capability coefficient}.}}
    \label{fig:SVB-overview}
\end{figure*}

We introduce \emph{subcircuit volumetric benchmarking} (SVB), a method for creating application-based benchmarks that \textit{do} track progress towards solving a given computational problem, e.g.~a utility-scale challenge problem. The SVB method, summarized in Fig.~\ref{fig:SVB-overview} and described in Sec.~\ref{sec:methods}, begins with one or more useful \textit{target} circuits that solve a relevant problem.  The target circuit[s] may be arbitrarily large.  Subcircuits of various sizes and shapes are “snipped out” of the target circuit[s], and incorporated into experiments run on a specific quantum computer to measure its ability to run the subcircuits without error.  The final output of an SVB experiment quantifies the largest subcircuits of the useful circuit[s] that the benchmarked quantum computer can reliably execute.  This approach enables benchmarking a quantum computer's ability to solve a particular computational problem even when the computer is far from being able to successfully solve the problem, e.g.~because its error rates are too high or it has too few qubits.  SVB thus enables fine-grained tracking of \textit{progress toward} the capability to solve any specific challenge problem.

SVB can also be used to predict how well the full circuit[s] would run on the tested system or on a larger but otherwise similar system, by extrapolating the process fidelity of the full circuit's execution from direct measurements of the process fidelity of small subcircuits of varied sizes.  We propose a simple formula for this extrapolation, based on ``effective error rates'', which we show to be accurate under strong assumptions and conjecture to be reliable in a wider range of practical scenarios.  It generalizes a simple and popular heuristic for predicting circuit fidelity---sum up the error rates (process infidelities) for each individual gate in the circuit \cite{Proctor2021-wt}, as (usually) estimated by randomized benchmarking \cite{Hashim2024-om, Emerson2005-fd, Emerson2007-am, Knill2008-jf, Magesan2011-hc, Proctor2019-gf}---but SVB extrapolation improves upon this simple heuristic by capturing commonly observed effects that the simple heuristic misses, including the effects of crosstalk \cite{Sarovar2020-pz,Proctor2021-wt, Gambetta2012-zd, Proctor2022-yl, Harper2023-pv} and constructive or destructive interference of coherent (unitary) errors. SVB therefore provides an elegant link between application-based benchmarks and the prediction of application performance from low-level metrics and models derived from (e.g.) randomized benchmarking or tomography \cite{Hashim2024-om}.

Application-based benchmarking is popular, in part, because it enables jointly testing an entire quantum computing ``stack'', including compilation algorithms \cite{Hines2023-be, Cross2019-ku, Proctor2025-cd, Hashim2024-om, Tomesh2022-nu, Lubinski2023-zy}. Until now, this kind of benchmarking was only possible for applications small enough to execute with reasonable fidelity on contemporary hardware \cite{Proctor2025-cd, Hashim2024-om, Lubinski2023-zy}, such as factoring 15 with Shor's algorithm. But an integrated quantum computer's performance on small, classically-tractable problems is not necessarily reflective of its performance on utility-scale problems \cite{Proctor2025-cd, Smolin2013-vc}, because (e.g.) unscalable circuit compilation strategies can be applied. SVB enables full-stack benchmarking with utility-scale applications even on current systems, by first compiling a utility-scale application and then using SVB to measure a system's capability on snippets from that circuit.

SVB can be applied to any quantum algorithm. In this paper, we demonstrate it by creating benchmarks from the Hamiltonian block-encoding subroutine \cite{low2017hamiltonian,chakraborty2018power,gilyen2019quantum} used in quantum chemistry applications, and running them on IBM Q systems (Sec.~\ref{sec:results}). Although these demonstrations did not use utility-scale problems (the largest ``application'' circuits we used are actually a subroutine for solving a classically easy problem in quantum chemistry, and comprise about 10,000 circuit layers on 21 qubits), they show clearly how SVB can be used to (1) track progress towards implementing utility-scale quantum algorithms and quantify it by a simple \emph{capability coefficient}, and (2) compare the performance of different algorithms or algorithm configurations that solve the same problem.

\section{Subcircuit Volumetric Benchmarking}\label{sec:methods}
Subcircuit volumetric benchmarking (SVB) is illustrated schematically in Fig.~\ref{fig:SVB-overview}.  It is designed to estimate how reliably a particular quantum computer can, or could, run a given quantum algorithm. The input to SVB is a quantum circuit $c$ that is \emph{fully compiled} for a quantum computer $\qc$ into gates intended to apply unitary operations.  By ``fully compiled'', we mean a circuit $c$ that contains only gates native to $\qc$, mapped to specific qubits in $\qc$ (see Appendix~\ref{app:circuits} for more details), as illustrated in the top row of Fig.~\ref{fig:SVB-overview}. SVB can be applied in principle to any circuit, but we envisage that $c$ will typically implement some interesting algorithm---perhaps a utility-scale algorithm requiring thousands of qubits and trillions of quantum gates (a.k.a.~\emph{quops})---or a subroutine from such an algorithm. Because SVB is designed to estimate the error in $\qc$'s execution of $c$, it does not directly quantify other aspects of a ``full stack'' quantum computing system like compiler performance or gate speed. But SVB can be used as part of a broader methodology for assessing full-stack performance, including algorithmic performance, e.g., by applying SVB to circuits created with different algorithmic parameters or created with different compilation algorithms.

Given a circuit $c$ for system $\qc$, an SVB experiment proceeds in two steps:
\begin{enumerate}
\item For each circuit shape $(w_i,d_i)$ in some set of user-chosen circuit shapes, sample $K$ subcircuits or ``snippets'' $c_{w_i,d_i,j}$ with that shape from $c$.
\item Experimentally estimate a metric for the quality with which each sampled subcircuit $c_{w_i,d_i,j}$ can be executed on $\qc$ using an efficient estimation protocol.
\end{enumerate}
Any procedure of this sort is an SVB experiment, but to design a specific SVB experiment it is necessary to choose a circuit snipping procedure, a quality metric, and a method for estimating that quality metric. In Section~\ref{subsec:subcirc-sel}, we explain precisely how we sample circuit snippets, and in Section~\ref{subsec:subcirc-fid-est} we explain the success metric we use---process fidelity---why we use that metric, and how we efficiently estimate it.

\begin{figure*}
    \centering
    \includegraphics[width=.95\textwidth]{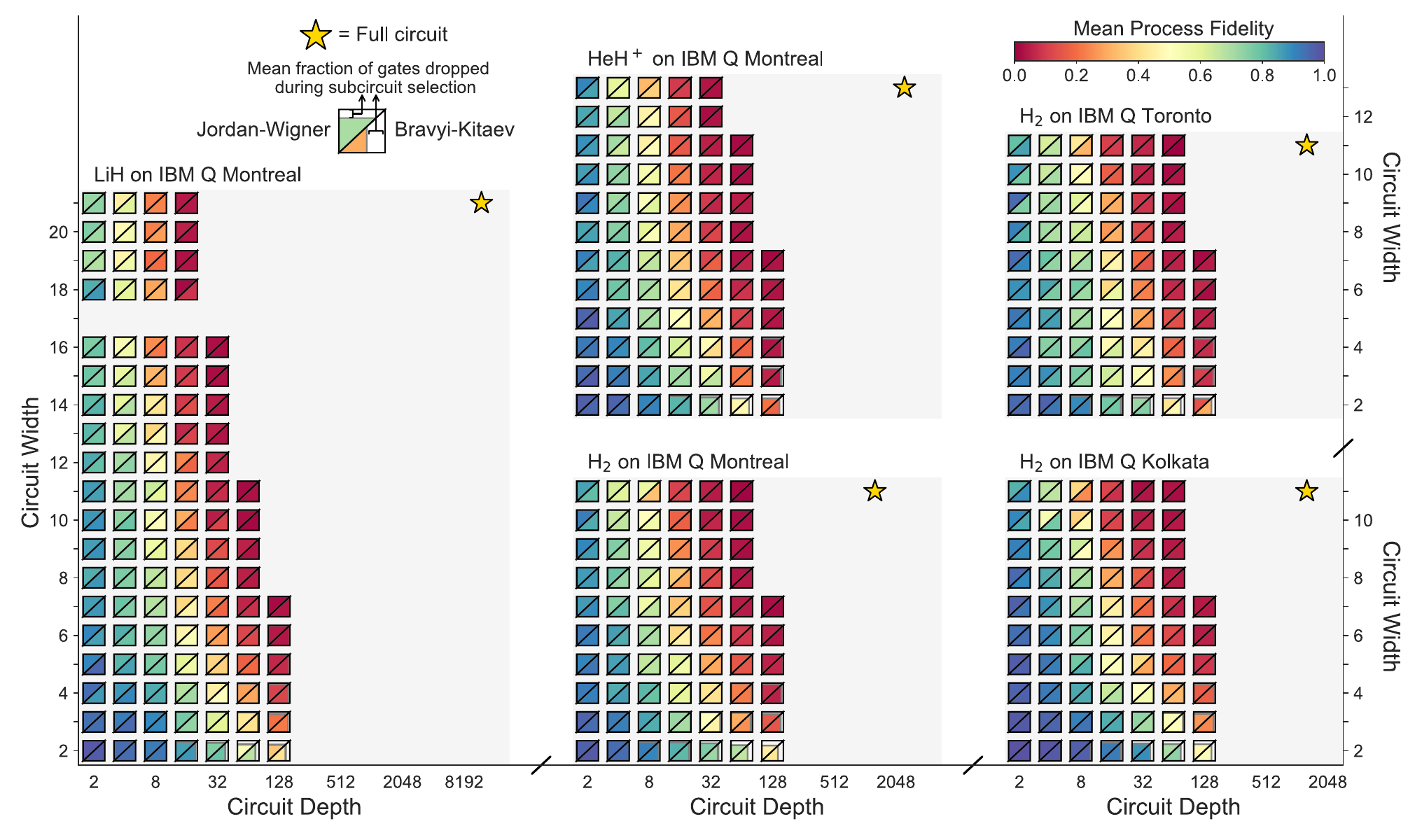}
    \caption{\small{\textbf{Benchmarking IBM Q system's performance on Hamiltonian simulation algorithms using SVB.} Volumetric benchmarks obtained by applying SVB to benchmark three IBM Q systems' performance on ground-state energy problems for three different molecules (LiH, HeH$^+$ and H$_2$) using two versions of a block-encoded Hamiltonian simulation algorithm. This data was collected in 2022. Each plot shows the measured performance of one system (IBM Q Montreal, IBM Q Toronto, or IBM Kolkata) on a ground-state energy algorithm for one molecule, using two different versions of the algorithm: a Jordan-Wigner (upper triangles) and Bravyi-Kitaev (lower triangles) encoding of the molecular orbitals into the qubits. In each case, we indicate the shape of the target circuit with a star in the upper right. In all cases, we observe little difference between the performance of the two versions of the algorithm. In all cases we also see that these systems were far from being capable of running the target circuit with low error (note the logarithmic scale on the horizontal axis).}}
    \label{fig:block-encoded-Ham-05-2022}
\end{figure*}

\begin{figure}
    \centering
    \includegraphics[width=.45\textwidth]{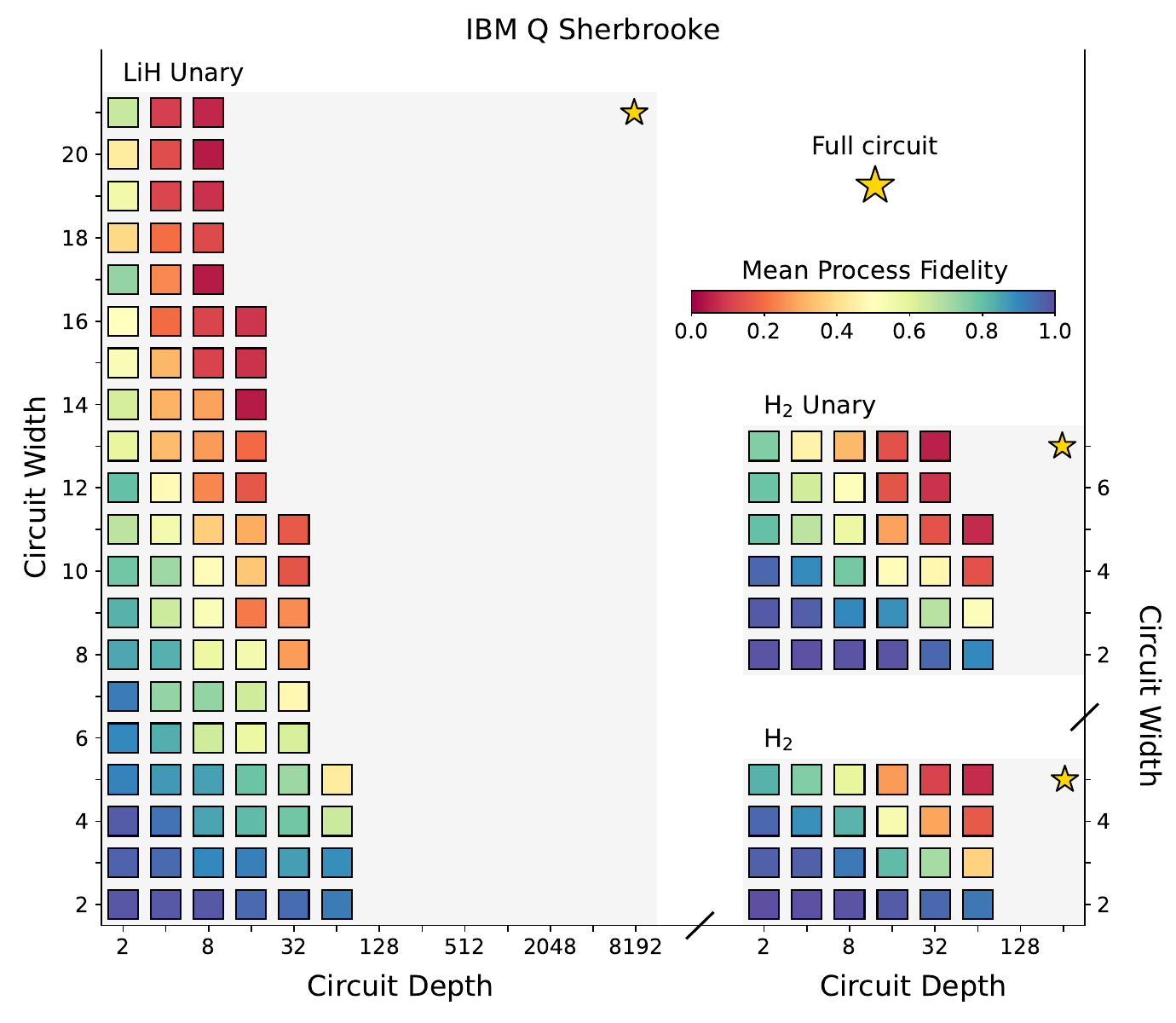}
    \caption{{\small\textbf{Benchmarking IBM Q Sherbrooke's performance on ground-state energy problems using SVB.} Volumetric benchmarks obtained by applying SVB to benchmark IBM Q Sherbrooke's performance on algorithms for ground-state energy problems: a block-encoded Hamiltonian simulation algorithm for LiH using a unary encoding, and for H$_2$ with and without a unary encoding. This data was collected in 2024. As in the IBM Q results from 2022 (Fig.~\ref{fig:block-encoded-Ham-05-2022}), IBM Q Sherbrooke is not capable of executing any of the target circuits with low error.}}
    \label{fig:block-encoded-Ham-sherbrooke}
\end{figure}

SVB data can be directly presented on a volumetric plot \cite{Proctor2021-wt,  Blume-Kohout2020-de} as illustrated in the lower right of Fig.~\ref{fig:SVB-overview} and demonstrated with experimental data in Figs.~\ref{fig:block-encoded-Ham-05-2022} and \ref{fig:block-encoded-Ham-sherbrooke}. SVB volumetric plots show (i) how the estimated quality metric (in our experiments, process fidelity) of circuit snippets depends on their shape $(w,d)$, and (ii) the shape of the full or ``target'' circuit $c$ (the yellow star in Fig.~\ref{fig:SVB-overview} and other figures). They illustrate both quantitatively and qualitatively how far the benchmarked quantum computer is from being able to run $c$ with a high probability of success. However, SVB data can be analyzed in greater detail to extract (e.g.) predictions for $c$'s fidelity and a capability coefficient, which we discuss later. Before doing so, we contrast SVB with existing approaches to application benchmarking.

\subsection{Features and advantages of SVB}
SVB is designed to overcome certain technical challenges to quantum computer benchmarking: efficiency, scalability, and the current unavailability of utility-scale quantum computers \cite{Proctor2025-cd}. These challenges are apparent in the simplest extant algorithm for estimating the error in a quantum computer's execution of a circuit $c$, which is to run $c$ many times on that quantum computer and compare the results to what \textit{would have} resulted from running $c$ on an ideal, error-free quantum computer. This approach is commonly used in existing quantum computer benchmarks \cite{Proctor2025-cd, Hashim2024-om, Lubinski2023-zy, Chen2022-dm, Tomesh2022-nu, Linke2017-mr, Wright2019-zj, Murali2019-my, Donkers2022-wt, Finzgar2022-aa, Mills2020-zh, Lubinski2023-zy, Lubinski2024-ci, Lubinski2023-mr, Chen2023-la, Benedetti2019-pp, Li2020-ry, Quetschlich2023-bg, Dong2021-gj, Martiel2021-vp, Van_der_Schoot2022-gv,Van_der_Schoot2023-vo, Cornelissen2021-yt, Georgopoulos2021-hh, Dong2022-ga, Chatterjee2025-lp}. To implement it, (i) a classical computer is used to compute or estimate the probability distribution over bit strings ($P_{\textrm{ideal}}$) from which $c$ would sample in the absence of errors, (ii) the circuit $c$ is executed many times on the quantum computer to estimate the distribution ($P_{\textrm{actual}}$) that is actually being sampled from, and (iii) the error in the execution in $c$ is quantified by computing a discrepancy metric $S$ between the \textit{estimated} $P_{\textrm{ideal}}$ and $P_{\textrm{actual}}$. An example of such a metric that is commonly used in benchmarking is the classical (Hellinger) fidelity \cite{Proctor2025-cd, Hashim2024-om, Lubinski2023-zy}. 

Unfortunately, that simple approach is generally intractable \cite{Proctor2025-cd, Hashim2024-om}. In general, the computational effort required to accurately estimate the quality metric $S$ grows superpolynomially with $c$'s width $w$ (the number of qubits).  Either or both of the quantum resources (number of circuit executions) or the classical resources (the amount of classical computing) required may blow up. Computing $P_{\textrm{ideal}}$ generally requires classical computations whose time and/or space requirements grow exponentially with $c$'s width. Similarly, a reasonable-precision estimate of the success metric $S$ (via estimating $P_{\textrm{actual}}$) generally requires a very large number of executions of the circuit $c$. 

Important exceptions exist. For example, a success metric can be directly estimated without first computing $P_{\textrm{ideal}}$ if $c$ solves a problem for which a candidate solution can be efficiently classified as correct or incorrect (such as factoring).  But for general circuits $c$, the simple approach remains intractable. Even when solutions can be efficiently classified, distinguishing very small success probabilities ($S$) from zero demands very many circuit executions. SVB avoids these problems; it neither computes  $P_{\textrm{ideal}}$ nor estimates $P_{\textrm{actual}}$ nor even requires running the target circuit $c$.

Another problem with simply running the target circuit $c$ is that it requires having access to a quantum computer $\qc$ large enough to run $c$.  One of the main reasons to benchmark contemporary quantum computers is to assess technological progress towards running ``useful'' utility-scale quantum algorithms \cite{Proctor2025-cd} (e.g., factoring RSA-2048), and these algorithms currently demand more qubits than are available in contemporary quantum computers. Today's quantum computers can be accurately regarded as \textit{prototypes} for future utility-scale quantum computers. When a user does not have $\qc$, but does have a prototype $\qc'$ that is similar but smaller (provides fewer qubits), then existing algorithm- and application-based benchmarks are limited to running smaller algorithmic circuits.  This only enables, at best, \textit{indirect} assessment of how far the prototype $\qc'$ is from successfully running a utility-scale circuit $c$.  SVB addresses this question directly by running ``snippets'' from the target circuit $c$ on the available prototype $\qc'$.

\subsection{Circuit Snipping}\label{subsec:subcirc-sel}
The central idea of SVB is to run circuits of various shapes that are snipped out from our target circuit $c$, which could be very large. There is, however, no unique way to do this and no natural \emph{and} precise set of criteria for judging between two different snipping algorithms---the best choice will depend on what is to be done with the SVB data (e.g., predicting the full circuit's fidelity). The primary factor that complicates how to do circuit snipping is multi-qubit gates: if we want to select a width-$w$ subcircuit from within a width-$w_c$ circuit $c$ and we randomly select $w$ qubits from $c$ to retain, there may be many gates between those qubits and qubits outside that set. If $w \ll w_c$ it will typically be the vast majority of those multi-qubit gates. For systems with a connectivity graph that is low-degree and local, there is a simple approach to solving this problem: pick \emph{connected} subsets of qubits, as shown in the schematic in the lower left of Fig.~\ref{fig:SVB-overview} (which shows a circuit for a linear chain of qubits), with the constructed circuit snippet then dropping any multi-qubit gates between those qubits and qubits outside that set. This is the approach we take, and below we describe it more precisely. Other approaches may be necessary for systems with high connectivity, and exploring snipping algorithms for that setting is left for future work.

Since SVB is intended to measure performance on circuits that may be too big to run on available hardware, we must consider adapting a circuit $c$ that was compiled for a hypothetical quantum computer $\qc$ with $n$ qubits to an available smaller prototype $\qc'$ that contains $n' < n$ qubits. We say that an $n'$-qubit subset of $\qc$ is \emph{equivalent} to $\qc'$ if $\qc'$ offers the same operations and connectivity graph as that subset, up to relabeling of qubits. To implement SVB on a prototype $\qc'$ for $\qc$, there must be at least one subset of $\qc$ that is equivalent to $\qc'$ and ideally there are many such subsets. For example, if $\qc$ is a $100 \times 100$ square grid of qubits and $\qc'$ is a $10 \times 10$ square grid of qubits, then there are 8281 different $10 \times 10$ grids of qubits embedded in $\qc$---each of which is equivalent to $\qc'$.

The input to our circuit snipping algorithm is a circuit $c$ compiled for $\qc$, with width $w_c$ (the number of qubits) and depth $d_c$ (the number of layers), a quantum computer $\qc'$ with $n'$ qubits on which we will execute our snippets, and a circuit shape $(w,d)$ for our snippets with $ w \leq n'$, $w \leq w_c$, and $d \leq d_c$. We assume that $c$ acts on all of the qubits in $\qc$ to simplify the description. With these inputs, we select a circuit snippet as follows:
\begin{enumerate}
    \itemsep0pt
    \item Uniformly sample a starting circuit layer from the first $d_c - d + 1$ layers in $c$, and create a depth $d$ circuit $c'$ from this and the following $d-1$ layers in $c$.
    \item Randomly sample an $n'$-qubit subset of $\qc$, denoted $q_n$, that is equivalent to $\qc'$.
    \item Randomly choose a set of $w$ qubits ($q_w$) from $q_n$ and which are connected in $\qc$ \footnote{That is the qubits in $q_w$ correspond to a connected sub-graph of $\qc$'s connectivity graph.}, and then create circuit $c_{w,d}$ of shape $(w,d)$  from $c'$ by discarding all the qubits in $c'$ except those in $q_w$ along with any multi-qubit gates between a qubit in $q_w$ and qubits outside of $q_w$ \footnote{We also relabel the qubits in our snippet, according to the equivalence mapping from our subset of $\qc$ to $\qc'$}.
\end{enumerate}
The random sampling in steps 2 and 3 is not necessarily \textit{uniform}, because uniform sampling over all valid qubit subsets may be both difficult (when $w \gg 1$) and suboptimal.  For example, sampling qubit subsets uniformly does not generally give each qubit equal probability of being included in a shape $(w,d)$ subcircuit. We use the following simple approach: to select $q_w$ we first select a single qubit from $c$ uniformly at random, and then we grow our set of selected qubits by randomly selecting a neighbour of our qubits to add to the set, until we have selected $w$ qubits.

Our snippet-sampling algorithm drops multi-qubit gates between qubit sets that cross the boundary of the selected subset. We tested the size of this effect---i.e., the fraction of the gates discarded by the algorithm---and found it to be small in the circuits used for this paper, except when $w=2$. This is shown in Fig.~\ref{fig:block-encoded-Ham-05-2022}, where the size of each triangle (the data points) shows the fraction of gates that were dropped.

\subsection{Estimating subcircuit performance}\label{subsec:subcirc-fid-est}
SVB estimates how some quality metric varies with circuit snippet shape, and so it is necessary to specify a quality metric and a method for estimating it. In this paper, we use \emph{process fidelity} \cite{Hashim2024-om}, because it can be efficiently estimated and plausibly extrapolated to larger circuits. In this subsection, we motivate the choice of process fidelity and explain how to estimate and extrapolate it.

\subsubsection{Definition and properties of process fidelity}
The process fidelity $F$ between an $n$-qubit unitary $U$ and a noisy implementation described by a transfer matrix $\Lambda$ is given by \cite{Hashim2024-om} 
\begin{equation}
    F = \bra{\Psi} \left(\mathbb{I} \otimes \mathcal{E} \right)[\ket{\Psi}{\bra{\Psi}}] \ket{\Psi},
\end{equation}
where $\mathcal{E} = \Lambda \mathcal{U}^{\dagger}$ is the circuit's error map (i.e., ideally $\mathcal{E}=\mathbb{I}$), $\mathcal{U}[\rho] = U\rho U^{\dagger}$ is the superoperator representation of the ideal unitary, $\mathbb{I}$ is the $n$-qubit identity superoperator, and $\ket{\Psi}$ is any pure state that is maximally entangled between the two $2^n$-dimensional Hilbert spaces on which $\mathbb{I} \otimes \mathcal{E}$ acts. Because process fidelity is a property of a dynamical map, it does not depend on or account for state preparation and measurement (SPAM) errors.

Process fidelity quantifies how accurately the quantum dynamical transformation $\Lambda$ (implemented by running $c$ on a real-world processor) simulates the ideal evolution $U$, if/when the qubits getting transformed are (1) initially entangled with other qubits and (2) not going to be measured immediately \cite{Hashim2024-om}. This is precisely the scenario that applies in SVB, because the snippets whose fidelity are being measured are subroutines of a larger target circuit.

\subsubsection{Predicting a circuit's process fidelity from subcircuit fidelities}
Using the process fidelity as a quality metric in SVB enables us to extract \emph{effective error rates} from SVB data, and then make principled predictions for the full circuit's process fidelity. We propose an extrapolation formula that predicts the full circuit's process fidelity, and we use it to define a \emph{capability coefficient} quantifying what fraction of the full circuit the benchmarked device can execute, as well as a \emph{scalability coefficient} that quantifies whether the device's performance degrades on larger circuits (and if it does, how much).

\begin{figure}
    \centering
    \includegraphics[width=.45\textwidth]{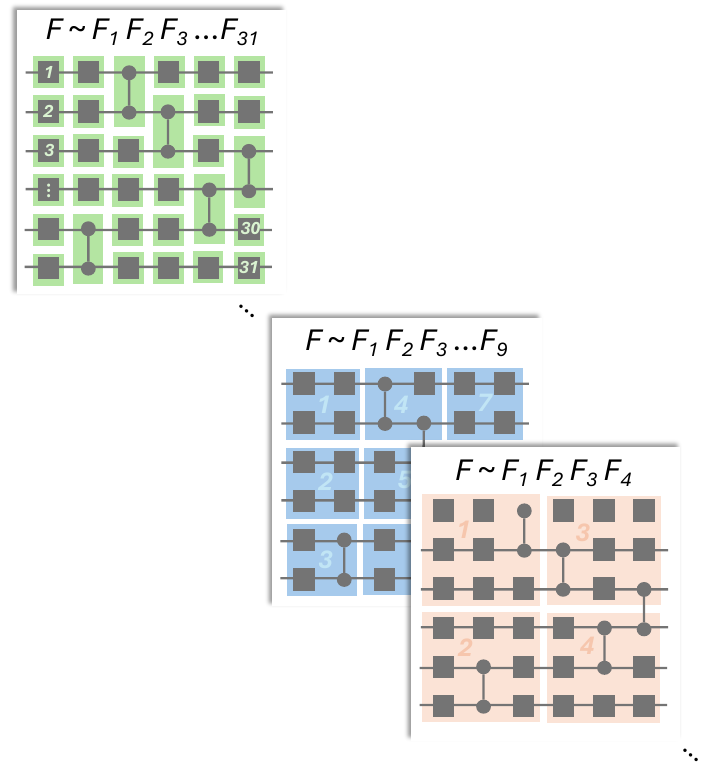}
    \caption{{\small\textbf{Estimating circuit fidelities from subcircuit fidelities.} The key idea underpinning SVB is that a large circuit's fidelity ($F$) can be estimated by estimating the fidelities of subcircuits from that circuit, then multiplying them together. We illustrate the idea by subdividing a circuit of shape $(6,6)$ into disjoint subcircuits of shape $(2,2)$ [middle] and $(3,3)$ [lower right], and estimating $F$ by the product of those subcircuits' fidelities ($F_i$). The fidelities of larger subcircuits can capture complex effects such as crosstalk and coherent addition or cancellation of error, which would not be captured by simply multiplying together the fidelities of all gates in the circuit (illustrated by the division of the same circuit into blocks of shape $(1,1)$ and $(2,1)$ in the top left). Two-qubit gates that connect different subcircuits cause practical and conceptual problems, as discussed in the main text. In SVB we randomly sample subcircuits of a given shape, rather than picking a disjoint subdivision of the circuit and exhaustively estimating the fidelities of all those circuits, as shown here.}}
    \label{fig:fidelity_schematic}
\end{figure}

Extrapolation is based on a simple principle: process fidelities are \emph{approximately multiplicative}. Consider a circuit $c$ of shape $(w_c, d_c)$ with a simple error model: local Pauli stochastic errors with a rate of $\epsilon$ and process fidelity $F = 1 - \epsilon$. For this error model, the fidelity of the circuit is given by \cite{Proctor2022-zs, Seth2025-zz}
\begin{equation}
    F_c \approx (1-\epsilon)^{w_cd_c} =  F^{w_cd_c},\label{eq:f_approx}
\end{equation}
to within, typically, only a small correction \footnote{There are two sources of approximation. First, there is an $\mathcal{O}(1/4^w)$ correction factor that is related to the difference between process fidelity and process polarization \cite{Hashim2024-om} (which, because we are typically interested in $w \gg 1$, we ignore for simplicity). Second, stochastic errors can cancel out within a circuit, and the exact rate of cancellation varies from circuit to circuit and depends on the exact biases in the errors. The impact of this effect on $F_{w,d}$ is at most $\mathcal{O}(\epsilon^2 w^2 d^2)$ but, in practice, it causes a negligible correction to Eq.~\eqref{eq:f_approx}.}. More generally, if each location $(i,j)$ in the circuit (i.e., the $i$th qubit at layer $j$) has local Pauli stochastic errors with error rate $\epsilon_{ij}$ then
\begin{equation}
    F_c \approx \prod_{ij} F_{ij}, \label{eq:f_approx2}
\end{equation}
where $F_{ij} = 1 - \epsilon_{ij}$.
This is the widely-used formula for approximating a circuit's fidelity by multiplying together the fidelities of its constituent gates, shown schematically in the top left of Fig.~\ref{fig:fidelity_schematic}.  It is often approximated further, using a first-order expansion, as $F_c = 1 - \sum_{ij}\epsilon_{ij}$.

For general errors, a circuit's fidelity is not well-approximated by Eq.~\eqref{eq:f_approx2}. Coherent errors on different gates can add or cancel, crosstalk errors can mean that a gate does not have a single, well-defined fidelity---a gate's errors can depend on the context of which other gates are being applied and may be correlated---and non-Markovian errors can complicate matters even further. This has been widely observed in experiments (e.g., see \cite{Proctor2021-wt}) and is one of the motivations for using SVB (or other holistic benchmarks) instead of simply measuring gate fidelities and using them to predict a circuit's fidelities. We can, however, generalize the ideas behind Eq.~\eqref{eq:f_approx2} to circuit snippets of arbitrary shape.

We can approximate $F_c$ by splitting $c$ into disjoint regions of any size, as shown in the diagram in Figure~\ref{fig:fidelity_schematic}. That is, the circuit is split into disjoint blocks each of shape $w \times d$ (or disjoint mixed-shape blocks) with $w \leq w_c$ qubits and $d \leq d_c$ layers, generalizing the $1 \times 1$ and $2 \times 1$ blocks used in Eq.~\eqref{eq:f_approx2} (shown in the top left of Fig.~\ref{fig:fidelity_schematic}). Letting $F^{(i)}_{w,d}$ be the fidelity of the $i$\textsuperscript{th} block, and letting $N_{w,d}$ denote the total number of blocks, then
\begin{equation}
    F_c \approx \prod_{i=1}^{N_{w,d}} F^{(i)}_{w,d}.\label{eq:f_approx3}
\end{equation}

As the size of the blocks grows, their error processes are, in typically circuits, likely to combine multiplicatively, as though they are stochastic. This follows from three general observations: (1) except in highly structured cases, the coherent errors in each large block's error process are unlikely to be close to maximally adding or cancelling between blocks (which requires those coherent errors to be commuting), meaning that the error processes approximately combine as though they are stochastic; (2) a growing fraction of crosstalk errors between nearby qubits will act \textit{within} each block rather than across blocks; and (3) a growing fraction of non-Markovian (time-correlated) errors will similarly act within each block and thus be captured by each block's process fidelity.  We therefore conjecture that as $w \to w_c$ and $d \to d_c$, the multiplicative approximation will become increasingly accurate.  (Note that when $(w,d) = (w_c,d_c)$ the equality holds trivially). If this is true, we can estimate $F_c$ by dividing it into smaller blocks and estimating each block's process fidelity. This conjecture is foundational to SVB.

\subsubsection{Predicting the target circuit's process fidelity with SVB data}
SVB does not actually divide $c$ into disjoint blocks and measure each one's fidelity.  For utility-scale circuits, the number of blocks would be prohibitive large.  Instead, it samples a relatively small number $K$ of snippets with each shape $(w,d)$, for some range of shapes, and estimates each snippet's process fidelity $F_{w,d,k}$ for $k=1,\dots,K$.  We then use the geometric mean of those snippets' observed process fidelity, in lieu of Eq.~\eqref{eq:f_approx3}, to estimate $\prod_{i=1}^{N_{w,d}} F^{(i)}_{w,d}$ and thus $F_c$. Specifically, we estimate $F_c$ as
\begin{align}
    \widehat{F}_c &= \textrm{GM}[F_{w,d}]^{\frac{w_c d_c}{wd}}\label{eq:Fc0}
\end{align}
where 
\begin{equation}
\textrm{GM}[F_{w,d}] \equiv \left( F_{w,d,1} F_{w,d,2} \cdots F_{w,d,K}\right)^{\frac{1}{K}}.
\end{equation}

Using the geometric mean fidelity of the snippets with shape $(w,d)$, we define an \emph{effective error per quop} in the context of $c$:
\begin{equation}
    \epsilon_{w,d} = 1 - \textrm{GM}[F_{w,d}]^{\frac{1}{wd}}. \label{eq:eepq}
\end{equation}
The estimate or prediction for $F_c$ derived from the shape $(w,d)$ snippets is thus
\begin{equation}
   \widehat{F}_c = (1 - \epsilon_{w,d})^{w_cd_c}.
\end{equation}

If the benchmarked system's errors are very well-behaved (e.g., local and gate-independent depolarizing errors), then the effective error rate will be nearly independent of snippet shape---there will be some $\epsilon$ for which $\epsilon_{w,d} \approx \epsilon$ for all $(w,d)$---and every snippet of the same shape will have approximately the same fidelity. Such a system has a single, well-defined error per quop $\epsilon$ in the context of circuit $c$, which can be easily extracted from SVB experiments. Generally, however, we expect the observed $\epsilon_{w,d}$ to vary with $(w,d)$, e.g.~because of crosstalk and/or coherent errors. SVB data can be used to probe (and perhaps understand) this dependence.

\subsubsection{Estimating process fidelity}
The SVB protocol is \textit{scalable} only if each step in the protocol can be performed with only polynomial resources even when $w \gg 1$.  Several methods for estimating a circuit snippet's process fidelity are both scalable and reliable (``system robust'' \cite{Proctor2025-cd}), each with different regimes of applicability \cite{Proctor2022-zs, Seth2025-zz, Hashim2024-om, Ferracin2021-vh}. We chose to use \emph{mirror circuit fidelity estimation} (MCFE) \cite{Proctor2022-zs}, which has broad applicability. MCFE estimates a circuit $c$'s process fidelity by running circuits sampled from three ensembles of \emph{mirror circuits} constructed from $c$. For the experiments in this paper, each circuit snippet was mapped to many different circuits (typically between a few hundred and a few thousand) that were run to estimate that snippet's fidelity. See Refs.~\cite{Proctor2022-zs, Hashim2024-om} for further details on MCFE.

Any experimental procedure for estimating a circuit's fidelity will result in uncertainty (``error bars'') in its estimate. Estimating the effective error per quop (Eq.~\eqref{eq:eepq}) requires estimating the geometric mean of $K$ circuits of the same shape $(w,d)$, and when the snippet's fidelities are close to zero the uncertainty on this estimate will be particularly large. In our analysis of experimental data below, we do not compute an effective error per quop at any circuit shape where we cannot clearly distinguish one or more of the estimated fidelities from zero (we use a somewhat arbitrary criteria of the fidelity estimate being below $7\%$, to account for both statistical uncertainty and systematic errors in MCFE estimates). 

\begin{figure*}
    \centering
    \includegraphics[width=.9\textwidth]{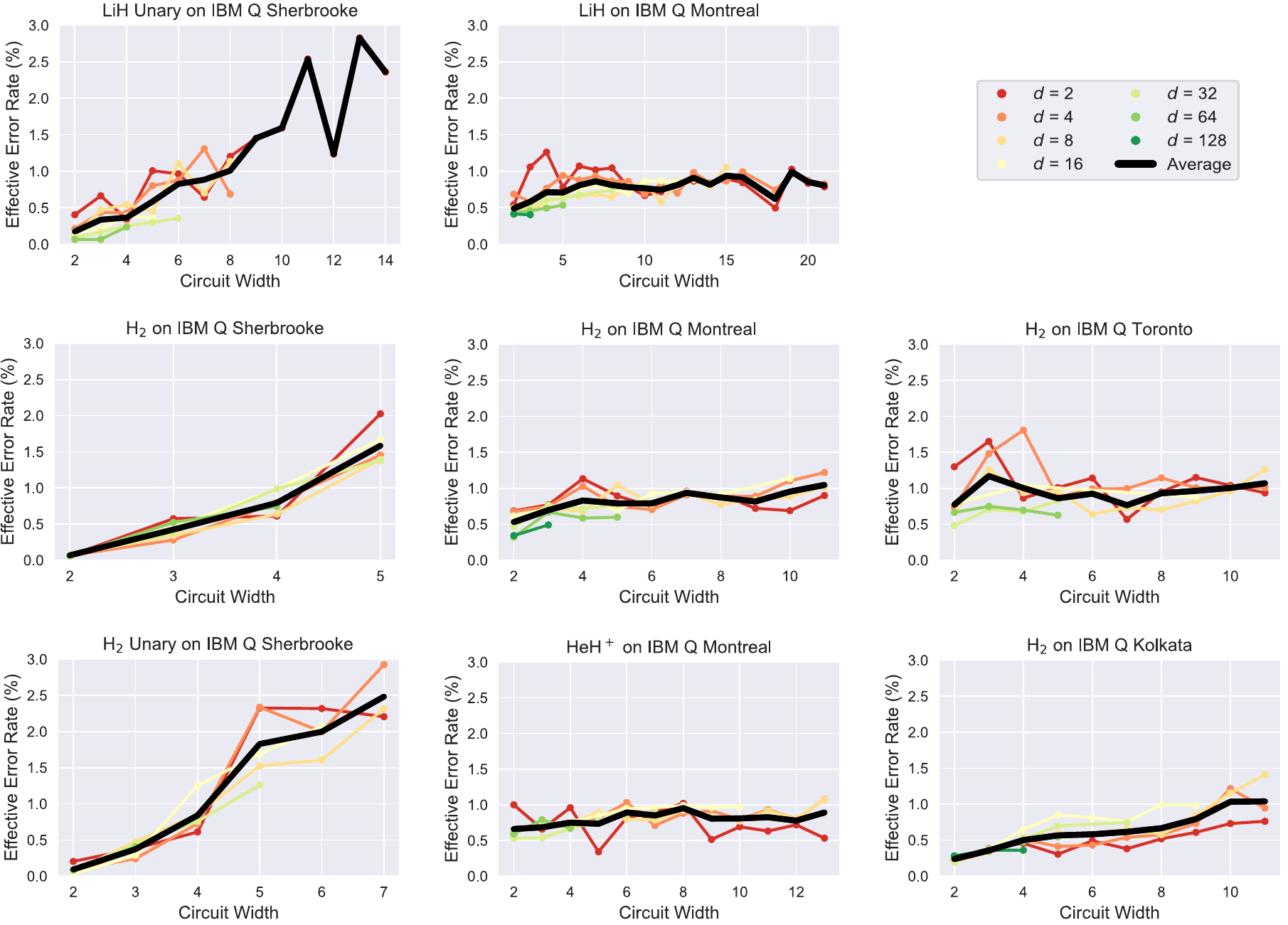}
    \caption{{\small\textbf{Effective error rates for IBM Q systems estimated from SVB.} The effective error rates observed in each SVB experiment (each panel) as a function of the circuit snippet's width and depth. In an SVB experiment we measure the process fidelities of circuit snippets of various shapes $(w,d)$, and for each $(w,d)$ we can turn those measured process fidelities into an effective error rate per quop $\epsilon_{w,d}$. We show $\epsilon_{w,d}$ as a function of width for each $d$ (colored points and lines) and the average over $d$ ($\epsilon_{w}$, black lines). We observe that $\epsilon_{w}$ grows with $w$, i.e., the effective error rate increases in circuits on more qubits, but its growth varies substantially between systems. In particular, although IBM Q Sherbrooke (left column) has a substantially lower effective error rate in the low-width ($w=2$) circuits than the other three systems (other columns), its effective error rate grows rapidly with width.}}
    \label{fig:effective_error_rates}
\end{figure*}

\section{Demonstration}\label{sec:results}
We now use SVB to create a scalable benchmark for a subroutine relevant to quantum chemistry calculations, and use it to benchmark IBM Q devices.

\subsection{The benchmarked Hamiltonian simulation subroutine: LCU circuits}
One prospective application of quantum computers is calculating the ground-state or low-lying excitations of molecular or solid-state systems~\cite{cao2019quantum, mcardle2020quantum}. A particular approach to this application involves sampling from the eigenspectrum of first-principles electronic structure Hamiltonians using quantum phase estimation~\cite{aspuru2005simulated}. This requires preparing specific encoded eigenstates of those Hamiltonians~\cite{pathak2023quantifying, berry2025rapid} and applying unitaries that encode their eigenspectra to those states~\cite{babbush2018encoding, russo2021evaluating}. Block encoding is a common subroutine that might be used for either purpose~\cite{babbush2018encoding, berry2025rapid}.  It is ubiquitous in contemporary quantum algorithms~\cite{gilyen2019quantum, martyn2021grand}. We demonstrate SVB for target circuits that implement a single application of the block encoding of a second-quantized electronic structure Hamiltonian, for different molecules.  While this is an algorithmic primitive (subroutine) rather than a full algorithm, it still yields target circuits too large to fit on present-day quantum computers.

We do not expect complete applications that use these block encodings to be feasible without fault-tolerant quantum computation (FTQC) facilitated by quantum error correction (QEC)~\cite{nelson2024assessment}. Even single applications of block encodings for minimal molecular Hamiltonians may not be feasible without QEC. Nevertheless, because these subroutines are key components of some of the most efficient quantum simulation algorithms~\cite{low2019hamiltonian,king2025quantum}, it is compelling to track progress towards their implementation as quantum hardware continues to mature. We leave the adaptation of SVB to benchmarking \textit{logical} circuits for future work.

The block encodings that we analyze with SVB are implemented using the \emph{linear combinations of unitaries} (LCU) method~\cite{childs2012hamiltonian}. We consider second-quantized Hamiltonians for three molecules: molecular hydrogen (H$_2$), the helium hydride cation (HeH$^+$), and lithium hydride (LiH). These Hamiltonians are discretized using small Gaussian basis sets: the STO-3G basis for H$_2$ and HeH$^+$, and the STO-6G basis with an active space of 4 orbitals for LiH \cite{hempel2018quantum, maupin2021variational}. The choice of molecules and basis sets means that all of our circuits use 21 or fewer qubits, although we note that SVB can be applied to circuits of any width. We also considered two different fermion-to-qubit mappings for encoding these Hamiltonians: Bravyi-Kitaev~\cite{bravyi2002fermionic} and Jordan-Wigner~\cite{jordan1928uber}. Further implementation details of the requisite LCU oracles are given in Appendix~\ref{app:demonstration-details}.

\subsection{LCU benchmarking experiments on IBM Q}
We ran SVB experiments using LCU circuits on various IBM Q systems, choosing a variety of combinations of the algorithmic parameters (see above) that define an LCU circuit instance. The precise details of how a high-level description of an LCU subroutine circuit was generated, optimized, compiled for a specific IBM Q system (creating our target $c$), and then turned into an SVB experiment, can be found in Appendix~\ref{app:demonstration-details}.

We conducted two separate sets of experiments, one in 2022 and the other in 2024. In our 2022 experiments, we ran SVBs for the three different molecules considered herein (H$_2$, HeH$^+$ and LiH) on the same system (IBM Q Montreal), and we ran an H$_2$ SVB experiment with identical algorithmic parameters on two other IBM Q systems (IBM Q Toronto and IBM Q Kolkata). In each case, we ran an experiment with the Bravyi-Kitaev encoding and one with the Jordan-Wigner encoding. In our 2024 experiments, we ran SVBs for H$_2$ and LiH on a single system (IBM Q Sherbrooke). In this experiment, we studied a smaller LCU circuit for H$_2$ by using a tapered Bravyi-Kitaev encoding \cite{bravyi2017tapering}. We also considered an alternative implementation of that LCU circuit using unary iteration~\cite{babbush2018encoding} (see Appendix~\ref{app:demonstration-details}).

\subsection{Summary volumetric plots}
Figures~\ref{fig:block-encoded-Ham-05-2022} and~\ref{fig:block-encoded-Ham-sherbrooke} summarize the results of these two experiments using volumetric plots. Each panel in each figure shows the results for a single device and LCU instance (labelled by the molecule and, in some cases, additional algorithmic parameters). In each panel, we show the shape of the target circuit (yellow stars)---i.e., the fully-compiled LCU circuit implementing the LCU unitary for that molecule. The data points (squares) show how the average process fidelities of the circuit snippets executed vary with their shape. In all cases, we find that the benchmarked systems are far from successfully implementing the target circuits---even for the example with smallest target circuit, corresponding to the H$_2$ molecule using a tapered Bravyi-Kitaev encoding (labelled as H$_2$ Tapered). These plots provide a qualitative summary of how far each system is from implementing the corresponding LCU circuit. We show how to quantify this ``distance'' to success in Section~\ref{sec:equops}. 

SVB enables experimental exploration of the impact of algorithmic parameters on circuit fidelities. Our first experiments were designed to explore one such parameter, the choice between Bravyi-Kitaev and Jordan-Wigner encodings. Figure~\ref{fig:block-encoded-Ham-05-2022} shows results for LCU circuits with a Bravyi-Kitaev encoding (lower triangles), and with a Jordan-Wigner encoding (upper triangles). For all systems, we observe only small differences between the process fidelities of the snippets for the Bravyi-Kitaev and Jordan-Wigner encodings. This is not surprising, but was also not a given; the target circuits defined by the two encodings have structural differences.  Because this analysis showed negligible differences between the two encodings, we combined the data from \textit{both} encodings for the subsequent analysis. This initial comparison demonstrates a general method for using SVB to compare a system's performance on different algorithms (or variants of the same algorithm) for the same problem. 

\begin{table}
\setlength{\tabcolsep}{5pt} 
\renewcommand{\arraystretch}{1.5} 
\begin{tabular}{ l  c  c  c  c  c}
Algorithm &	System	&  Width & Depth & $F_0$ & $F$ \\
\toprule
H$_2$ 	& Toronto & 11 & 1277 & $10^{-47}$ & $10^{-66}$ \\ 
H$_2$ 	& Kolkata & 11 & 1277 & $10^{-14}$ & $10^{-64}$ \\ 
H$_2$ 	& Montreal & 11	& 1277 & $10^{-32}$ & $10^{-64}$ \\ 
HeH$^+$ & Montreal	& 13 & 2480	& $10^{-93}$ & $10^{-125}$ \\ 
LiH	& Montreal & 21 & 12090	& $10^{-542}$ & $10^{-894}$ \\
LiH	& Sherbrooke & 21 & 7979 & $10^{-127}$ & $10^{-1738}$ \\
H$_2$ Tapered & Sherbrooke & 5 & 263 & $10^{-0.4}$ & $10^{-9}$ \\ 
H$_2$ Unary  & Sherbrooke &7 & 252 & $10^{-0.7}$ & $10^{-19}$ \\
\end{tabular}
\caption{{\small\textbf{Predicted target circuit fidelities}. For each algorithm and system that we benchmarked, we use the measured fidelities of circuit snippets to predict the fidelity with which that system can run the target algorithmic circuit $c$. When we use the fidelities of the narrowest circuits ($w=2$) to predicted $c$'s fidelity we obtain the ``optimistic'' estimates $F_0$, which we expect will typically be similar to predictions obtained from measuring one- and two-qubit gate errors rates with RB. When we instead use the measured fidelities of wide snippets (in all cases here, as wide as the target circuit) we obtain fidelity predictions $F$ that we conjecture will be much more predictive of $c$'s process fidelity. In all cases, we see that $F$ is many orders of magnitude smaller than $F_0$, and in the cases of H$_2$ on Sherbrooke we observe $F_0 > 0.1$ whereas $F$ is so small that it would require an infeasible number of circuit executions to obtain useable data.}}\label{tab:error_rates}
\end{table}

\begin{table*}[ht!]
\setlength{\tabcolsep}{3pt} 
\renewcommand{\arraystretch}{1.7} 
\begin{tabular}{ l  c  c  c  c | c  c}
Algorithm &	System	&  Algorithm Size  & Predicted Capability  & Observed Capability &  Scalability Coefficient & \textbf{Capbility Coefficient} \\
 &		&  (quops, $Q_T$) &  (quops, $Q_0$) &  (quops, $Q_C$) &  ($Q_C/Q_0$, $\%$) & \textbf{($Q_C/Q_T$, $\%$)}  \\
\toprule
 H$_2$ 	& Toronto & 14000 & 130 & 94  & 72\% &  \textbf{0.7\%} \\ 
 H$_2$ 	& Kolkata & 14000 & 420 & 96  & 23\% &  \textbf{0.7\%} \\ 
 H$_2$ 	& Montreal & 14000 & 190 & 96  & 50\% &  \textbf{0.7\%} \\ 
 HeH$^+$ & Montreal	& 32000 & 150 & 112  & 74\% &  \textbf{0.3\%} \\ 
 LiH	&  Montreal & 254000 & 203 & 124  & 60\% &  \textbf{0.05\%} \\
LiH Unary	& Sherbrooke & 	169000 & 570 & 42 & 7\%  &  \textbf{0.03\%} \\
H$_2$ Tapered & Sherbrooke & 1300 & 1480 & 63   &	4\% &  \textbf{5\%} \\ 
H$_2$ Tapered Unary  & Sherbrooke & 1800 & 1100 & 40 & 4\%  &  \textbf{2\%} \\
\end{tabular}
\caption{{\small\textbf{Summarizing system performance with observed quops.} The observed capability ($Q_C$) in quops, of each system on each LCU instance, from which we can concisely summarize how close the system is to successfully running the target circuit by computing the \emph{capability coefficient} (right hand column), which is the observed quops as a percentage of the number of quops in the target circuit ($Q_T$). The observed capability is computed from the effective error rates (Fig.~\ref{fig:effective_error_rates}) observed in each SVB experiment, and we compare it to the capability predicted from the lowest width circuits ($Q_0$). The \textit{scalability coefficient} $Q_C/Q_0$ quantifies the effect of complex errors (e.g., crosstalk) in degrading system performance on wide circuits.}}\label{tab:quops}
\end{table*}

\subsection{Summarizing performance with effective quops}\label{sec:equops}

We now show how to use SVB results to concisely quantify how close a system is to implementing an algorithm, demonstrated with these example SVB experiments. At each snippet shape $(w,d)$ we compute the effective error per quop ($\epsilon_{w,d}$) using Eq.~\eqref{eq:eepq}. These error rates enable us to predict the target circuit fidelity and quantify how far the benchmarked device is from being able to execute the target circuit, as a percentage of that circuit's quops. 

Figure \ref{fig:effective_error_rates} shows how $\epsilon_{w,d}$ varies with snippet width ($w$) for each experiment, showing both the average over depths ($\epsilon_{w}$, black lines) and $\epsilon_{w,d}$ for each depth $d$ (colored lines). The dependence of the effective error per quop on width reveals both spatial variation in qubit error rates and crosstalk errors in that system. In all cases, we find that $\epsilon_{w}$ grows with $w$, but its growth varies substantially from system to system. It increases by an order of magnitude with all three SVBs run on IBM Q Sherbrooke, from around $0.2 \%$ at $w=2$ to between around $1.5\%$ and around $2\%$ at $w_{\max}$ ($w_{\max}$ denotes the largest width, which varies between algorithms, and is, in our experiments, equal to the target circuit's width). This dependence of $\epsilon_{w}$ on $w$ shows that context-independent gate error rates (as, e.g., measured using RB) will not be able to accurately predict the target circuit's fidelity $F_c$. We conjecture that SVB will enable more accurate predictions of $F_c$ by measuring the error per quop in a context that closely mimics execution of the target circuit.

To quantify the effect of the context-dependent error rates, we predicted $F_c$ from the observed effective error rates at the lowest width ($w=2$) and at the largest width ($w = w_{\max}$). We denote these predictions by $F_0$ and $F$ respectively. They are shown in Table~\ref{tab:error_rates} for all our experiments. In all cases, $F$ is smaller than $F_0$ by orders of magnitude (the smallest discrepancy is approximately eight orders of magnitude). For the smallest target circuit (H$_2$ Tapered) $F_0 = 0.40 \approx 10^{-0.4}$ but $F = 10^{-9}$. The LCU circuit is a subroutine in useful quantum algorithms, where it is typically used many times, so even a fidelity of $0.40$ is far too small to obtain reasonable fidelities in a Hamiltonian simulation algorithm. However, an $\mathcal{O}(1)$ fidelity for a complete algorithmic circuit might be sufficient to obtain utility, e.g., with the application of error mitigation \cite{Kim2023-si}. In contrast, a fidelity of $\mathcal{O}(10^{-9})$ is unlikely to be consist with obtaining utility. This highlights the importance of realistic estimates of a target circuit's fidelity, as enabled by SVB. 

To quantify how far a system is from successfully running the target circuits, we can express our effective error per quops as an \emph{effective quops},
\begin{equation}
    Q_{w,d} = 1/\epsilon_{w,d}.
\end{equation}
This is the maximum number of operations that can be implemented while achieving a fidelity of $\mathcal{O}(1)$. The effective quops could depend strongly on both snippet width and depth, but we conjecture that \textit{in practice} $\epsilon_{w,d}$ will asymptote to a constant value $\epsilon_c$ for sufficiently large $w$ and $d$ (see Section~\ref{subsec:subcirc-fid-est})---and that ``sufficiently large'' widths and depths will be much smaller than the target circuit's width and depth, except for target circuits where the width or depth is small. If this conjecture holds, a single \emph{effective quops in context} can be defined for a target circuit $c$ as 
\begin{equation}
Q_{c} = 1/\epsilon_{c}.
\end{equation}
Our experiments did not probe sufficiently large $w$ and $d$ to test this conjecture, but we can examine how $Q_{w,d}$ varied with shape in our experiments. We observed relatively little dependence on $d$ (see Fig.~\ref{fig:effective_error_rates}, noting that we do not compute $\epsilon_{w,d}$ when we cannot distinguish one or more fidelities from zero), but there is significant dependence on $w$.  We can take advantage of the relatively small variance with respect to $d$ to define $Q_{w} = 1/\epsilon_{w}$, where $\epsilon_{w}$ is the average of $\epsilon_{w,d}$ over all depths $d$ (black lines in Fig.~\ref{fig:effective_error_rates}). Since $Q_w$ depends strongly on $w$ in most of the experiments we performed, we define each system's \emph{observed capability} as $Q_{C} = 1/\epsilon_{w_{\max}}$.

It is instructive to compare each system's observed capability to the capability that would be predicted from running only circuits of width $w=2$, which we call \emph{predicted capability}:  $Q_{0} = 1/\epsilon_2$. We expect $Q_0$ to be similar to what RB error rates would predict, except that $Q_0$ is more context-specific because it is extracted from narrow snippets of the target circuit instead of random Clifford circuits.  Table~\ref{tab:quops} tabulates the observed and predicted capabilities for each experiment.  Observed capabilities range from $Q_{C} \sim 40$ up to $Q_{C} \sim 120$. In almost all cases they are much smaller than the corresponding predicted capability. We quantify this gap with the \emph{scalability coefficient}: $Q_C/Q_0$, shown in Table~\ref{tab:error_rates} (as a percentage). The scalability coefficient ranges from 74\% down to 4\%.

Finally, we use our results to quantify how close these systems are to accurately executing the target circuits. Each target circuit requires a certain number of quops, $Q_T = w_cd_c$ \footnote{We do not distinguish between idle operations and gates, because, in real systems, idle operations are often as noisy as gates and cannot be \emph{a priori} assumed to have very low error.}. Since we have measured the number of quops that each system can execute (in context), we can concisely summarize how close each system is to executing the target circuit with the ratio $Q_C/Q_T$, which we call the \emph{capability coefficient}. These systems achieve between $0.03\%$ and $5\%$ of the required quops needed to implement these LCU circuits. The capability coefficient provides a simple way to track a sequence of prototypes' performance on an algorithm even when that algorithm can only be run with vanishing chance of success. The smallest capability coefficient we observed ($0.03\%$) corresponds to executing the target circuit with a fidelity of around $10^{-1700}$ (see Table~\ref{tab:error_rates}), which could not feasibly be measured by just running that target circuit.

\section{Conclusion}\label{sec:conclusion}
Quantum computer benchmarks can enable measuring progress towards quantum utility, but only if they are scalable, reliable, and measure well-motivated metrics. In this work, we have introduced subcircuit volumetric benchmarking (SVB), which is a method for creating benchmarks that meet these criteria out of any quantum algorithm or computational problem. The central idea of SVB is to run subcircuits snipped out from a potentially very large circuit that implements an algorithm or algorithmic subroutine of interest. By snipping out circuits of various shapes, we can measure how complex and arbitrary kinds of errors like crosstalk, coherent, or non-Markovian errors impact execution of the target circuit---even on systems that are far too small or too noisy to run the target circuit with reasonable fidelity.  SVB data enables summarizing performance using effective error rates, and can even be used to define an \emph{observed quops} that can be compared directly to the size of the target circuit. This simple idea enables predicting the fidelity with which a larger or better (but otherwise comparable) quantum computer would run the target circuit. How close \textit{this} system is to successfully running the target circuit can be concisely summarized by the capability coefficient, which is the observed quops divided by the number of quops in the target circuit (and, herein, expressed as a percentage).

Quantum utility will likely require FTQC architectures \cite{Campbell2017-tw} to obtain the large number of quops (low error rates) that appear necessary for solving practical problems with quantum algorithms \cite{Proctor2025-cd,Rubin2024-gc, low2025fast, gidney2025factor}. We demonstrated SVB with circuits run directly on physical qubits---known as ``NISQ computing''---but SVB can equally well be applied to circuits run on fault-tolerant logical qubits protected by error correction. SVB could be used as part of a framework for comparing NISQ computing and FTQC solutions to the same computational problem: a NISQ compilation and an FTQC compilation for a problem could be defined, the effective quops of a system (running in NISQ and FTQC modes) measured for each compilation, and the capability coefficient computed for each. However, FTQC and NISQ compilations of algorithms are \textit{very} different \cite{Campbell2017-tw} (e.g., because universal gates in FTQC require techniques such as magic state distillation or cultivation) and the error processes that will occur on logical qubits in real FTQC systems remain to be seen. 

\begin{acknowledgments}
{\small
We gratefully acknowledge useful conversations with Anand Ganti, Lucas Kovalsky, Andrew Landahl, Alicia Magann, Benjamin Morrison, Jake Nelson, Corey Ostrove, Shivesh Pathak, Mohan Sarovar, and Noah Siekierski. This material was funded in part by the U.S. Department of Energy, Office of Science, Office of Advanced Scientific Computing Research Quantum Testbed Pathfinder Program and in part by the U.S. Department of Energy, National Nuclear Security Administration, Advanced Simulation and Computing program. S.K.S. and A.D.B. acknowledge support from the Department of Energy (DOE) Office of Fusion Energy Sciences ``Foundations for quantum simulation of warm dense matter'' project. T.P. acknowledges support from an Office of Advanced Scientific Computing Research Early Career Award. Sandia National Laboratories is a multi-program laboratory managed and operated by National Technology and Engineering Solutions of Sandia, LLC., a wholly owned subsidiary of Honeywell International, Inc., for the U.S. Department of Energy's National Nuclear Security Administration under contract DE-NA-0003525. This research used IBM Quantum resources of the Air Force Research Laboratory.  All statements of fact, opinion or conclusions contained herein are those of the authors and should not be construed as representing the official views or policies of the U.S. Department of Energy, or the U.S. Government, or IBM, or the IBM Quantum team.}
\end{acknowledgments}

\bibliography{bibliography}

\begin{thebibliography}{83}%
\makeatletter
\providecommand \@ifxundefined [1]{%
 \@ifx{#1\undefined}
}%
\providecommand \@ifnum [1]{%
 \ifnum #1\expandafter \@firstoftwo
 \else \expandafter \@secondoftwo
 \fi
}%
\providecommand \@ifx [1]{%
 \ifx #1\expandafter \@firstoftwo
 \else \expandafter \@secondoftwo
 \fi
}%
\providecommand \natexlab [1]{#1}%
\providecommand \enquote  [1]{``#1''}%
\providecommand \bibnamefont  [1]{#1}%
\providecommand \bibfnamefont [1]{#1}%
\providecommand \citenamefont [1]{#1}%
\providecommand \href@noop [0]{\@secondoftwo}%
\providecommand \href [0]{\begingroup \@sanitize@url \@href}%
\providecommand \@href[1]{\@@startlink{#1}\@@href}%
\providecommand \@@href[1]{\endgroup#1\@@endlink}%
\providecommand \@sanitize@url [0]{\catcode `\\12\catcode `\$12\catcode
  `\&12\catcode `\#12\catcode `\^12\catcode `\_12\catcode `\%12\relax}%
\providecommand \@@startlink[1]{}%
\providecommand \@@endlink[0]{}%
\providecommand \url  [0]{\begingroup\@sanitize@url \@url }%
\providecommand \@url [1]{\endgroup\@href {#1}{\urlprefix }}%
\providecommand \urlprefix  [0]{URL }%
\providecommand \Eprint [0]{\href }%
\providecommand \doibase [0]{https://doi.org/}%
\providecommand \selectlanguage [0]{\@gobble}%
\providecommand \bibinfo  [0]{\@secondoftwo}%
\providecommand \bibfield  [0]{\@secondoftwo}%
\providecommand \translation [1]{[#1]}%
\providecommand \BibitemOpen [0]{}%
\providecommand \bibitemStop [0]{}%
\providecommand \bibitemNoStop [0]{.\EOS\space}%
\providecommand \EOS [0]{\spacefactor3000\relax}%
\providecommand \BibitemShut  [1]{\csname bibitem#1\endcsname}%
\let\auto@bib@innerbib\@empty
\bibitem [{\citenamefont {Proctor}\ \emph {et~al.}(2025)\citenamefont
  {Proctor}, \citenamefont {Young}, \citenamefont {Baczewski},\ and\
  \citenamefont {Blume-Kohout}}]{Proctor2025-cd}%
  \BibitemOpen
  \bibfield  {author} {\bibinfo {author} {\bibfnamefont {T.}~\bibnamefont
  {Proctor}}, \bibinfo {author} {\bibfnamefont {K.}~\bibnamefont {Young}},
  \bibinfo {author} {\bibfnamefont {A.~D.}\ \bibnamefont {Baczewski}},\ and\
  \bibinfo {author} {\bibfnamefont {R.}~\bibnamefont {Blume-Kohout}},\
  }\bibfield  {title} {\bibinfo {title} {Benchmarking quantum computers},\
  }\href {https://doi.org/10.1038/s42254-024-00796-z} {\bibfield  {journal}
  {\bibinfo  {journal} {Nat. Rev. Phys.}\ }\textbf {\bibinfo {volume} {7}},\
  \bibinfo {pages} {105} (\bibinfo {year} {2025})}\BibitemShut {NoStop}%
\bibitem [{\citenamefont {Hashim}\ \emph {et~al.}(2024)\citenamefont {Hashim},
  \citenamefont {Nguyen}, \citenamefont {Goss}, \citenamefont {Marinelli},
  \citenamefont {Naik}, \citenamefont {Chistolini}, \citenamefont {Hines},
  \citenamefont {Marceaux}, \citenamefont {Kim}, \citenamefont {Gokhale},
  \citenamefont {Tomesh}, \citenamefont {Chen}, \citenamefont {Jiang},
  \citenamefont {Ferracin}, \citenamefont {Rudinger}, \citenamefont {Proctor},
  \citenamefont {Young}, \citenamefont {Blume-Kohout},\ and\ \citenamefont
  {Siddiqi}}]{Hashim2024-om}%
  \BibitemOpen
  \bibfield  {author} {\bibinfo {author} {\bibfnamefont {A.}~\bibnamefont
  {Hashim}}, \bibinfo {author} {\bibfnamefont {L.~B.}\ \bibnamefont {Nguyen}},
  \bibinfo {author} {\bibfnamefont {N.}~\bibnamefont {Goss}}, \bibinfo {author}
  {\bibfnamefont {B.}~\bibnamefont {Marinelli}}, \bibinfo {author}
  {\bibfnamefont {R.~K.}\ \bibnamefont {Naik}}, \bibinfo {author}
  {\bibfnamefont {T.}~\bibnamefont {Chistolini}}, \bibinfo {author}
  {\bibfnamefont {J.}~\bibnamefont {Hines}}, \bibinfo {author} {\bibfnamefont
  {J.~P.}\ \bibnamefont {Marceaux}}, \bibinfo {author} {\bibfnamefont
  {Y.}~\bibnamefont {Kim}}, \bibinfo {author} {\bibfnamefont {P.}~\bibnamefont
  {Gokhale}}, \bibinfo {author} {\bibfnamefont {T.}~\bibnamefont {Tomesh}},
  \bibinfo {author} {\bibfnamefont {S.}~\bibnamefont {Chen}}, \bibinfo {author}
  {\bibfnamefont {L.}~\bibnamefont {Jiang}}, \bibinfo {author} {\bibfnamefont
  {S.}~\bibnamefont {Ferracin}}, \bibinfo {author} {\bibfnamefont
  {K.}~\bibnamefont {Rudinger}}, \bibinfo {author} {\bibfnamefont
  {T.}~\bibnamefont {Proctor}}, \bibinfo {author} {\bibfnamefont {K.~C.}\
  \bibnamefont {Young}}, \bibinfo {author} {\bibfnamefont {R.}~\bibnamefont
  {Blume-Kohout}},\ and\ \bibinfo {author} {\bibfnamefont {I.}~\bibnamefont
  {Siddiqi}},\ }\bibfield  {title} {\bibinfo {title} {A practical introduction
  to benchmarking and characterization of quantum computers},\ }\href
  {https://arxiv.org/abs/2408.12064} {\bibfield  {journal} {\bibinfo  {journal}
  {arXiv [quant-ph]}\ } (\bibinfo {year} {2024})},\ \Eprint
  {https://arxiv.org/abs/2408.12064} {arXiv:2408.12064 [quant-ph]} \BibitemShut
  {NoStop}%
\bibitem [{\citenamefont {Lubinski}\ \emph
  {et~al.}(2023{\natexlab{a}})\citenamefont {Lubinski}, \citenamefont {Johri},
  \citenamefont {Varosy}, \citenamefont {Coleman}, \citenamefont {Zhao},
  \citenamefont {Necaise}, \citenamefont {Baldwin}, \citenamefont {Mayer},\
  and\ \citenamefont {Proctor}}]{Lubinski2023-zy}%
  \BibitemOpen
  \bibfield  {author} {\bibinfo {author} {\bibfnamefont {T.}~\bibnamefont
  {Lubinski}}, \bibinfo {author} {\bibfnamefont {S.}~\bibnamefont {Johri}},
  \bibinfo {author} {\bibfnamefont {P.}~\bibnamefont {Varosy}}, \bibinfo
  {author} {\bibfnamefont {J.}~\bibnamefont {Coleman}}, \bibinfo {author}
  {\bibfnamefont {L.}~\bibnamefont {Zhao}}, \bibinfo {author} {\bibfnamefont
  {J.}~\bibnamefont {Necaise}}, \bibinfo {author} {\bibfnamefont {C.~H.}\
  \bibnamefont {Baldwin}}, \bibinfo {author} {\bibfnamefont {K.}~\bibnamefont
  {Mayer}},\ and\ \bibinfo {author} {\bibfnamefont {T.}~\bibnamefont
  {Proctor}},\ }\bibfield  {title} {\bibinfo {title} {{Application-Oriented}
  performance benchmarks for quantum computing},\ }\href
  {https://doi.org/10.1109/TQE.2023.3253761} {\bibfield  {journal} {\bibinfo
  {journal} {IEEE Transactions on Quantum Engineering}\ }\textbf {\bibinfo
  {volume} {4}},\ \bibinfo {pages} {1} (\bibinfo {year}
  {2023}{\natexlab{a}})}\BibitemShut {NoStop}%
\bibitem [{\citenamefont {Amico}\ \emph {et~al.}(2023)\citenamefont {Amico},
  \citenamefont {Zhang}, \citenamefont {Jurcevic}, \citenamefont {Bishop},
  \citenamefont {Nation}, \citenamefont {Wack},\ and\ \citenamefont
  {McKay}}]{Amico2023-ze}%
  \BibitemOpen
  \bibfield  {author} {\bibinfo {author} {\bibfnamefont {M.}~\bibnamefont
  {Amico}}, \bibinfo {author} {\bibfnamefont {H.}~\bibnamefont {Zhang}},
  \bibinfo {author} {\bibfnamefont {P.}~\bibnamefont {Jurcevic}}, \bibinfo
  {author} {\bibfnamefont {L.~S.}\ \bibnamefont {Bishop}}, \bibinfo {author}
  {\bibfnamefont {P.}~\bibnamefont {Nation}}, \bibinfo {author} {\bibfnamefont
  {A.}~\bibnamefont {Wack}},\ and\ \bibinfo {author} {\bibfnamefont {D.~C.}\
  \bibnamefont {McKay}},\ }\bibfield  {title} {\bibinfo {title} {Defining
  standard strategies for quantum benchmarks},\ }\href@noop {} {\bibfield
  {journal} {\bibinfo  {journal} {arXiv preprint arXiv:2303.02108}\ } (\bibinfo
  {year} {2023})}\BibitemShut {NoStop}%
\bibitem [{\citenamefont {Proctor}\ \emph {et~al.}(2021)\citenamefont
  {Proctor}, \citenamefont {Rudinger}, \citenamefont {Young}, \citenamefont
  {Nielsen},\ and\ \citenamefont {Blume-Kohout}}]{Proctor2021-wt}%
  \BibitemOpen
  \bibfield  {author} {\bibinfo {author} {\bibfnamefont {T.}~\bibnamefont
  {Proctor}}, \bibinfo {author} {\bibfnamefont {K.}~\bibnamefont {Rudinger}},
  \bibinfo {author} {\bibfnamefont {K.}~\bibnamefont {Young}}, \bibinfo
  {author} {\bibfnamefont {E.}~\bibnamefont {Nielsen}},\ and\ \bibinfo {author}
  {\bibfnamefont {R.}~\bibnamefont {Blume-Kohout}},\ }\bibfield  {title}
  {\bibinfo {title} {Measuring the capabilities of quantum computers},\ }\href
  {https://doi.org/10.1038/s41567-021-01409-7} {\bibfield  {journal} {\bibinfo
  {journal} {Nat. Phys.}\ }\textbf {\bibinfo {volume} {18}},\ \bibinfo {pages}
  {75} (\bibinfo {year} {2021})}\BibitemShut {NoStop}%
\bibitem [{\citenamefont {Blume-Kohout}\ and\ \citenamefont
  {Young}(2020)}]{Blume-Kohout2020-de}%
  \BibitemOpen
  \bibfield  {author} {\bibinfo {author} {\bibfnamefont {R.}~\bibnamefont
  {Blume-Kohout}}\ and\ \bibinfo {author} {\bibfnamefont {K.~C.}\ \bibnamefont
  {Young}},\ }\bibfield  {title} {\bibinfo {title} {A volumetric framework for
  quantum computer benchmarks},\ }\href
  {https://doi.org/10.22331/q-2020-11-15-362} {\bibfield  {journal} {\bibinfo
  {journal} {Quantum}\ }\textbf {\bibinfo {volume} {4}},\ \bibinfo {pages}
  {362} (\bibinfo {year} {2020})}\BibitemShut {NoStop}%
\bibitem [{\citenamefont {Erhard}\ \emph {et~al.}(2019)\citenamefont {Erhard},
  \citenamefont {Wallman}, \citenamefont {Postler}, \citenamefont {Meth},
  \citenamefont {Stricker}, \citenamefont {Martinez}, \citenamefont
  {Schindler}, \citenamefont {Monz}, \citenamefont {Emerson},\ and\
  \citenamefont {Blatt}}]{Erhard2019-wk}%
  \BibitemOpen
  \bibfield  {author} {\bibinfo {author} {\bibfnamefont {A.}~\bibnamefont
  {Erhard}}, \bibinfo {author} {\bibfnamefont {J.~J.}\ \bibnamefont {Wallman}},
  \bibinfo {author} {\bibfnamefont {L.}~\bibnamefont {Postler}}, \bibinfo
  {author} {\bibfnamefont {M.}~\bibnamefont {Meth}}, \bibinfo {author}
  {\bibfnamefont {R.}~\bibnamefont {Stricker}}, \bibinfo {author}
  {\bibfnamefont {E.~A.}\ \bibnamefont {Martinez}}, \bibinfo {author}
  {\bibfnamefont {P.}~\bibnamefont {Schindler}}, \bibinfo {author}
  {\bibfnamefont {T.}~\bibnamefont {Monz}}, \bibinfo {author} {\bibfnamefont
  {J.}~\bibnamefont {Emerson}},\ and\ \bibinfo {author} {\bibfnamefont
  {R.}~\bibnamefont {Blatt}},\ }\bibfield  {title} {\bibinfo {title}
  {Characterizing large-scale quantum computers via cycle benchmarking},\
  }\href {https://doi.org/10.1038/s41467-019-13068-7} {\bibfield  {journal}
  {\bibinfo  {journal} {Nat. Commun.}\ }\textbf {\bibinfo {volume} {10}},\
  \bibinfo {pages} {5347} (\bibinfo {year} {2019})}\BibitemShut {NoStop}%
\bibitem [{\citenamefont {Cross}\ \emph {et~al.}(2019)\citenamefont {Cross},
  \citenamefont {Bishop}, \citenamefont {Sheldon}, \citenamefont {Nation},\
  and\ \citenamefont {Gambetta}}]{Cross2019-ku}%
  \BibitemOpen
  \bibfield  {author} {\bibinfo {author} {\bibfnamefont {A.~W.}\ \bibnamefont
  {Cross}}, \bibinfo {author} {\bibfnamefont {L.~S.}\ \bibnamefont {Bishop}},
  \bibinfo {author} {\bibfnamefont {S.}~\bibnamefont {Sheldon}}, \bibinfo
  {author} {\bibfnamefont {P.~D.}\ \bibnamefont {Nation}},\ and\ \bibinfo
  {author} {\bibfnamefont {J.~M.}\ \bibnamefont {Gambetta}},\ }\bibfield
  {title} {\bibinfo {title} {Validating quantum computers using randomized
  model circuits},\ }\href {https://doi.org/10.1103/PhysRevA.100.032328}
  {\bibfield  {journal} {\bibinfo  {journal} {Phys. Rev. A}\ }\textbf {\bibinfo
  {volume} {100}},\ \bibinfo {pages} {032328} (\bibinfo {year}
  {2019})}\BibitemShut {NoStop}%
\bibitem [{\citenamefont {Chen}\ \emph {et~al.}(2022)\citenamefont {Chen},
  \citenamefont {Fang}, \citenamefont {Guan}, \citenamefont {Hong},
  \citenamefont {Huang}, \citenamefont {Liu}, \citenamefont {Wang},\ and\
  \citenamefont {Ying}}]{Chen2022-dm}%
  \BibitemOpen
  \bibfield  {author} {\bibinfo {author} {\bibfnamefont {K.}~\bibnamefont
  {Chen}}, \bibinfo {author} {\bibfnamefont {W.}~\bibnamefont {Fang}}, \bibinfo
  {author} {\bibfnamefont {J.}~\bibnamefont {Guan}}, \bibinfo {author}
  {\bibfnamefont {X.}~\bibnamefont {Hong}}, \bibinfo {author} {\bibfnamefont
  {M.}~\bibnamefont {Huang}}, \bibinfo {author} {\bibfnamefont
  {J.}~\bibnamefont {Liu}}, \bibinfo {author} {\bibfnamefont {Q.}~\bibnamefont
  {Wang}},\ and\ \bibinfo {author} {\bibfnamefont {M.}~\bibnamefont {Ying}},\
  }\bibfield  {title} {\bibinfo {title} {{VeriQBench}: A benchmark for multiple
  types of quantum circuits},\ }\bibfield  {journal} {\bibinfo  {journal}
  {arXiv preprint arXiv:2206.10880}\ }\href
  {https://doi.org/1048550/arXiv.2206.10880} {1048550/arXiv.2206.10880}
  (\bibinfo {year} {2022})\BibitemShut {NoStop}%
\bibitem [{\citenamefont {{Tomesh}}\ \emph {et~al.}(2022)\citenamefont
  {{Tomesh}}, \citenamefont {{Gokhale}}, \citenamefont {{Omole}}, \citenamefont
  {{Ravi}}, \citenamefont {{Smith}}, \citenamefont {{Viszlai}}, \citenamefont
  {{Wu}}, \citenamefont {{Hardavellas}}, \citenamefont {{Martonosi}},\ and\
  \citenamefont {{Chong}}}]{Tomesh2022-nu}%
  \BibitemOpen
  \bibfield  {author} {\bibinfo {author} {\bibnamefont {{Tomesh}}}, \bibinfo
  {author} {\bibnamefont {{Gokhale}}}, \bibinfo {author} {\bibnamefont
  {{Omole}}}, \bibinfo {author} {\bibnamefont {{Ravi}}}, \bibinfo {author}
  {\bibnamefont {{Smith}}}, \bibinfo {author} {\bibnamefont {{Viszlai}}},
  \bibinfo {author} {\bibnamefont {{Wu}}}, \bibinfo {author} {\bibnamefont
  {{Hardavellas}}}, \bibinfo {author} {\bibnamefont {{Martonosi}}},\ and\
  \bibinfo {author} {\bibnamefont {{Chong}}},\ }\bibfield  {title} {\bibinfo
  {title} {{SupermarQ}: A scalable quantum benchmark suite},\ }in\ \href
  {https://doi.org/10.1109/HPCA53966.2022.00050} {\emph {\bibinfo {booktitle}
  {2022 {IEEE} International Symposium on {High-Performance} Computer
  Architecture ({HPCA})}}},\ Vol.~\bibinfo {volume} {0}\ (\bibinfo {year}
  {2022})\ pp.\ \bibinfo {pages} {587--603}\BibitemShut {NoStop}%
\bibitem [{\citenamefont {Linke}\ \emph {et~al.}(2017)\citenamefont {Linke},
  \citenamefont {Maslov}, \citenamefont {Roetteler}, \citenamefont {Debnath},
  \citenamefont {Figgatt}, \citenamefont {Landsman}, \citenamefont {Wright},\
  and\ \citenamefont {Monroe}}]{Linke2017-mr}%
  \BibitemOpen
  \bibfield  {author} {\bibinfo {author} {\bibfnamefont {N.~M.}\ \bibnamefont
  {Linke}}, \bibinfo {author} {\bibfnamefont {D.}~\bibnamefont {Maslov}},
  \bibinfo {author} {\bibfnamefont {M.}~\bibnamefont {Roetteler}}, \bibinfo
  {author} {\bibfnamefont {S.}~\bibnamefont {Debnath}}, \bibinfo {author}
  {\bibfnamefont {C.}~\bibnamefont {Figgatt}}, \bibinfo {author} {\bibfnamefont
  {K.~A.}\ \bibnamefont {Landsman}}, \bibinfo {author} {\bibfnamefont
  {K.}~\bibnamefont {Wright}},\ and\ \bibinfo {author} {\bibfnamefont
  {C.}~\bibnamefont {Monroe}},\ }\bibfield  {title} {\bibinfo {title}
  {Experimental comparison of two quantum computing architectures},\ }\href
  {https://doi.org/10.1073/pnas.1618020114} {\bibfield  {journal} {\bibinfo
  {journal} {Proc. Natl. Acad. Sci. U. S. A.}\ }\textbf {\bibinfo {volume}
  {114}},\ \bibinfo {pages} {3305} (\bibinfo {year} {2017})}\BibitemShut
  {NoStop}%
\bibitem [{\citenamefont {Wright}\ \emph {et~al.}(2019)\citenamefont {Wright},
  \citenamefont {Beck}, \citenamefont {Debnath}, \citenamefont {Amini},
  \citenamefont {Nam}, \citenamefont {Grzesiak}, \citenamefont {Chen},
  \citenamefont {Pisenti}, \citenamefont {Chmielewski}, \citenamefont
  {Collins}, \citenamefont {Hudek}, \citenamefont {Mizrahi}, \citenamefont
  {Wong-Campos}, \citenamefont {Allen}, \citenamefont {Apisdorf}, \citenamefont
  {Solomon}, \citenamefont {Williams}, \citenamefont {Ducore}, \citenamefont
  {Blinov}, \citenamefont {Kreikemeier}, \citenamefont {Chaplin}, \citenamefont
  {Keesan}, \citenamefont {Monroe},\ and\ \citenamefont {Kim}}]{Wright2019-zj}%
  \BibitemOpen
  \bibfield  {author} {\bibinfo {author} {\bibfnamefont {K.}~\bibnamefont
  {Wright}}, \bibinfo {author} {\bibfnamefont {K.~M.}\ \bibnamefont {Beck}},
  \bibinfo {author} {\bibfnamefont {S.}~\bibnamefont {Debnath}}, \bibinfo
  {author} {\bibfnamefont {J.~M.}\ \bibnamefont {Amini}}, \bibinfo {author}
  {\bibfnamefont {Y.}~\bibnamefont {Nam}}, \bibinfo {author} {\bibfnamefont
  {N.}~\bibnamefont {Grzesiak}}, \bibinfo {author} {\bibfnamefont {J.-S.}\
  \bibnamefont {Chen}}, \bibinfo {author} {\bibfnamefont {N.~C.}\ \bibnamefont
  {Pisenti}}, \bibinfo {author} {\bibfnamefont {M.}~\bibnamefont
  {Chmielewski}}, \bibinfo {author} {\bibfnamefont {C.}~\bibnamefont
  {Collins}}, \bibinfo {author} {\bibfnamefont {K.~M.}\ \bibnamefont {Hudek}},
  \bibinfo {author} {\bibfnamefont {J.}~\bibnamefont {Mizrahi}}, \bibinfo
  {author} {\bibfnamefont {J.~D.}\ \bibnamefont {Wong-Campos}}, \bibinfo
  {author} {\bibfnamefont {S.}~\bibnamefont {Allen}}, \bibinfo {author}
  {\bibfnamefont {J.}~\bibnamefont {Apisdorf}}, \bibinfo {author}
  {\bibfnamefont {P.}~\bibnamefont {Solomon}}, \bibinfo {author} {\bibfnamefont
  {M.}~\bibnamefont {Williams}}, \bibinfo {author} {\bibfnamefont {A.~M.}\
  \bibnamefont {Ducore}}, \bibinfo {author} {\bibfnamefont {A.}~\bibnamefont
  {Blinov}}, \bibinfo {author} {\bibfnamefont {S.~M.}\ \bibnamefont
  {Kreikemeier}}, \bibinfo {author} {\bibfnamefont {V.}~\bibnamefont
  {Chaplin}}, \bibinfo {author} {\bibfnamefont {M.}~\bibnamefont {Keesan}},
  \bibinfo {author} {\bibfnamefont {C.}~\bibnamefont {Monroe}},\ and\ \bibinfo
  {author} {\bibfnamefont {J.}~\bibnamefont {Kim}},\ }\bibfield  {title}
  {\bibinfo {title} {Benchmarking an 11-qubit quantum computer},\ }\href
  {https://doi.org/10.1038/s41467-019-13534-2} {\bibfield  {journal} {\bibinfo
  {journal} {Nat. Commun.}\ }\textbf {\bibinfo {volume} {10}},\ \bibinfo
  {pages} {5464} (\bibinfo {year} {2019})}\BibitemShut {NoStop}%
\bibitem [{\citenamefont {Murali}\ \emph {et~al.}(2019)\citenamefont {Murali},
  \citenamefont {Linke}, \citenamefont {Martonosi}, \citenamefont {Abhari},
  \citenamefont {Nguyen},\ and\ \citenamefont {Alderete}}]{Murali2019-my}%
  \BibitemOpen
  \bibfield  {author} {\bibinfo {author} {\bibfnamefont {P.}~\bibnamefont
  {Murali}}, \bibinfo {author} {\bibfnamefont {N.~M.}\ \bibnamefont {Linke}},
  \bibinfo {author} {\bibfnamefont {M.}~\bibnamefont {Martonosi}}, \bibinfo
  {author} {\bibfnamefont {A.~J.}\ \bibnamefont {Abhari}}, \bibinfo {author}
  {\bibfnamefont {N.~H.}\ \bibnamefont {Nguyen}},\ and\ \bibinfo {author}
  {\bibfnamefont {C.~H.}\ \bibnamefont {Alderete}},\ }\bibfield  {title}
  {\bibinfo {title} {Full-stack, real-system quantum computer studies:
  architectural comparisons and design insights},\ }in\ \href
  {https://doi.org/10.1145/3307650.3322273} {\emph {\bibinfo {booktitle}
  {Proceedings of the 46th International Symposium on Computer
  Architecture}}},\ \bibinfo {series and number} {ISCA '19}\ (\bibinfo
  {publisher} {Association for Computing Machinery},\ \bibinfo {address} {New
  York, NY, USA},\ \bibinfo {year} {2019})\ pp.\ \bibinfo {pages}
  {527--540}\BibitemShut {NoStop}%
\bibitem [{\citenamefont {Donkers}\ \emph {et~al.}(2022)\citenamefont
  {Donkers}, \citenamefont {Mesman}, \citenamefont {Al-Ars},\ and\
  \citenamefont {M{\"o}ller}}]{Donkers2022-wt}%
  \BibitemOpen
  \bibfield  {author} {\bibinfo {author} {\bibfnamefont {H.}~\bibnamefont
  {Donkers}}, \bibinfo {author} {\bibfnamefont {K.}~\bibnamefont {Mesman}},
  \bibinfo {author} {\bibfnamefont {Z.}~\bibnamefont {Al-Ars}},\ and\ \bibinfo
  {author} {\bibfnamefont {M.}~\bibnamefont {M{\"o}ller}},\ }\bibfield  {title}
  {\bibinfo {title} {{QPack} scores: Quantitative performance metrics for
  application-oriented quantum computer benchmarking},\ }\bibfield  {journal}
  {\bibinfo  {journal} {arXiv preprint arXiv:2205.12142}\ }\href
  {https://doi.org/1048550/arXiv.2205.12142} {1048550/arXiv.2205.12142}
  (\bibinfo {year} {2022})\BibitemShut {NoStop}%
\bibitem [{\citenamefont {Fin{\v z}gar}\ \emph {et~al.}(2022)\citenamefont
  {Fin{\v z}gar}, \citenamefont {Ross}, \citenamefont {Klepsch},\ and\
  \citenamefont {Luckow}}]{Finzgar2022-aa}%
  \BibitemOpen
  \bibfield  {author} {\bibinfo {author} {\bibfnamefont {J.~R.}\ \bibnamefont
  {Fin{\v z}gar}}, \bibinfo {author} {\bibfnamefont {P.}~\bibnamefont {Ross}},
  \bibinfo {author} {\bibfnamefont {J.}~\bibnamefont {Klepsch}},\ and\ \bibinfo
  {author} {\bibfnamefont {A.}~\bibnamefont {Luckow}},\ }\bibfield  {title}
  {\bibinfo {title} {{QUARK}: A framework for quantum computing application
  benchmarking},\ }\bibfield  {journal} {\bibinfo  {journal} {arXiv preprint
  arXiv:2202.03028}\ }\href {https://doi.org/1048550/arXiv.2202.03028}
  {1048550/arXiv.2202.03028} (\bibinfo {year} {2022})\BibitemShut {NoStop}%
\bibitem [{\citenamefont {Mills}\ \emph {et~al.}(2020)\citenamefont {Mills},
  \citenamefont {Sivarajah}, \citenamefont {Scholten},\ and\ \citenamefont
  {Duncan}}]{Mills2020-zh}%
  \BibitemOpen
  \bibfield  {author} {\bibinfo {author} {\bibfnamefont {D.}~\bibnamefont
  {Mills}}, \bibinfo {author} {\bibfnamefont {S.}~\bibnamefont {Sivarajah}},
  \bibinfo {author} {\bibfnamefont {T.~L.}\ \bibnamefont {Scholten}},\ and\
  \bibinfo {author} {\bibfnamefont {R.}~\bibnamefont {Duncan}},\ }\bibfield
  {title} {\bibinfo {title} {{Application-Motivated}, holistic benchmarking of
  a full quantum computing stack},\ }\bibfield  {journal} {\bibinfo  {journal}
  {arXiv preprint arXiv:2006.01273}\ }\href
  {https://doi.org/1048550/arXiv.2006.01273} {1048550/arXiv.2006.01273}
  (\bibinfo {year} {2020})\BibitemShut {NoStop}%
\bibitem [{\citenamefont {Lubinski}\ \emph {et~al.}(2024)\citenamefont
  {Lubinski}, \citenamefont {Goings}, \citenamefont {Mayer}, \citenamefont
  {Johri}, \citenamefont {Reddy}, \citenamefont {Mehta}, \citenamefont
  {Bhatia}, \citenamefont {Rappaport}, \citenamefont {Mills}, \citenamefont
  {Baldwin}, \citenamefont {Zhao}, \citenamefont {Barbosa}, \citenamefont
  {Maity},\ and\ \citenamefont {Mundada}}]{Lubinski2024-ci}%
  \BibitemOpen
  \bibfield  {author} {\bibinfo {author} {\bibfnamefont {T.}~\bibnamefont
  {Lubinski}}, \bibinfo {author} {\bibfnamefont {J.~J.}\ \bibnamefont
  {Goings}}, \bibinfo {author} {\bibfnamefont {K.}~\bibnamefont {Mayer}},
  \bibinfo {author} {\bibfnamefont {S.}~\bibnamefont {Johri}}, \bibinfo
  {author} {\bibfnamefont {N.}~\bibnamefont {Reddy}}, \bibinfo {author}
  {\bibfnamefont {A.}~\bibnamefont {Mehta}}, \bibinfo {author} {\bibfnamefont
  {N.}~\bibnamefont {Bhatia}}, \bibinfo {author} {\bibfnamefont
  {S.}~\bibnamefont {Rappaport}}, \bibinfo {author} {\bibfnamefont
  {D.}~\bibnamefont {Mills}}, \bibinfo {author} {\bibfnamefont {C.~H.}\
  \bibnamefont {Baldwin}}, \bibinfo {author} {\bibfnamefont {L.}~\bibnamefont
  {Zhao}}, \bibinfo {author} {\bibfnamefont {A.}~\bibnamefont {Barbosa}},
  \bibinfo {author} {\bibfnamefont {S.}~\bibnamefont {Maity}},\ and\ \bibinfo
  {author} {\bibfnamefont {P.~S.}\ \bibnamefont {Mundada}},\ }\bibfield
  {title} {\bibinfo {title} {Quantum algorithm exploration using
  {Application-Oriented} performance benchmarks},\ }\bibfield  {journal}
  {\bibinfo  {journal} {arXiv preprint arXiv:2402.08985}\ }\href
  {https://doi.org/1048550/arXiv.2402.08985} {1048550/arXiv.2402.08985}
  (\bibinfo {year} {2024})\BibitemShut {NoStop}%
\bibitem [{\citenamefont {Lubinski}\ \emph
  {et~al.}(2023{\natexlab{b}})\citenamefont {Lubinski}, \citenamefont
  {Coffrin}, \citenamefont {McGeoch}, \citenamefont {Sathe}, \citenamefont
  {Apanavicius},\ and\ \citenamefont {Bernal~Neira}}]{Lubinski2023-mr}%
  \BibitemOpen
  \bibfield  {author} {\bibinfo {author} {\bibfnamefont {T.}~\bibnamefont
  {Lubinski}}, \bibinfo {author} {\bibfnamefont {C.}~\bibnamefont {Coffrin}},
  \bibinfo {author} {\bibfnamefont {C.}~\bibnamefont {McGeoch}}, \bibinfo
  {author} {\bibfnamefont {P.}~\bibnamefont {Sathe}}, \bibinfo {author}
  {\bibfnamefont {J.}~\bibnamefont {Apanavicius}},\ and\ \bibinfo {author}
  {\bibfnamefont {D.~E.}\ \bibnamefont {Bernal~Neira}},\ }\bibfield  {title}
  {\bibinfo {title} {Optimization applications as quantum performance
  benchmarks},\ }\bibfield  {journal} {\bibinfo  {journal} {arXiv preprint
  arXiv:2302.02278}\ }\href {https://doi.org/1048550/arXiv.2302.02278}
  {1048550/arXiv.2302.02278} (\bibinfo {year} {2023}{\natexlab{b}})\BibitemShut
  {NoStop}%
\bibitem [{\citenamefont {Chen}\ \emph {et~al.}(2023)\citenamefont {Chen},
  \citenamefont {Nielsen}, \citenamefont {Ebert}, \citenamefont {Inlek},
  \citenamefont {Wright}, \citenamefont {Chaplin}, \citenamefont {Maksymov},
  \citenamefont {P{\'a}ez}, \citenamefont {Poudel}, \citenamefont {Maunz},\
  and\ \citenamefont {Gamble}}]{Chen2023-la}%
  \BibitemOpen
  \bibfield  {author} {\bibinfo {author} {\bibfnamefont {J.-S.}\ \bibnamefont
  {Chen}}, \bibinfo {author} {\bibfnamefont {E.}~\bibnamefont {Nielsen}},
  \bibinfo {author} {\bibfnamefont {M.}~\bibnamefont {Ebert}}, \bibinfo
  {author} {\bibfnamefont {V.}~\bibnamefont {Inlek}}, \bibinfo {author}
  {\bibfnamefont {K.}~\bibnamefont {Wright}}, \bibinfo {author} {\bibfnamefont
  {V.}~\bibnamefont {Chaplin}}, \bibinfo {author} {\bibfnamefont
  {A.}~\bibnamefont {Maksymov}}, \bibinfo {author} {\bibfnamefont
  {E.}~\bibnamefont {P{\'a}ez}}, \bibinfo {author} {\bibfnamefont
  {A.}~\bibnamefont {Poudel}}, \bibinfo {author} {\bibfnamefont
  {P.}~\bibnamefont {Maunz}},\ and\ \bibinfo {author} {\bibfnamefont
  {J.}~\bibnamefont {Gamble}},\ }\bibfield  {title} {\bibinfo {title}
  {Benchmarking a trapped-ion quantum computer with 29 algorithmic qubits},\
  }\bibfield  {journal} {\bibinfo  {journal} {arXiv preprint arXiv:2308.05071}\
  }\href {https://doi.org/1048550/arXiv.2308.05071} {1048550/arXiv.2308.05071}
  (\bibinfo {year} {2023})\BibitemShut {NoStop}%
\bibitem [{\citenamefont {Benedetti}\ \emph {et~al.}(2019)\citenamefont
  {Benedetti}, \citenamefont {Garcia-Pintos}, \citenamefont {Perdomo},
  \citenamefont {Leyton-Ortega}, \citenamefont {Nam},\ and\ \citenamefont
  {Perdomo-Ortiz}}]{Benedetti2019-pp}%
  \BibitemOpen
  \bibfield  {author} {\bibinfo {author} {\bibfnamefont {M.}~\bibnamefont
  {Benedetti}}, \bibinfo {author} {\bibfnamefont {D.}~\bibnamefont
  {Garcia-Pintos}}, \bibinfo {author} {\bibfnamefont {O.}~\bibnamefont
  {Perdomo}}, \bibinfo {author} {\bibfnamefont {V.}~\bibnamefont
  {Leyton-Ortega}}, \bibinfo {author} {\bibfnamefont {Y.}~\bibnamefont {Nam}},\
  and\ \bibinfo {author} {\bibfnamefont {A.}~\bibnamefont {Perdomo-Ortiz}},\
  }\bibfield  {title} {\bibinfo {title} {A generative modeling approach for
  benchmarking and training shallow quantum circuits},\ }\href
  {https://doi.org/10.1038/s41534-019-0157-8} {\bibfield  {journal} {\bibinfo
  {journal} {npj Quantum Information}\ }\textbf {\bibinfo {volume} {5}},\
  \bibinfo {pages} {45} (\bibinfo {year} {2019})}\BibitemShut {NoStop}%
\bibitem [{\citenamefont {Li}\ and\ \citenamefont
  {Krishnamoorthy}(2020)}]{Li2020-ry}%
  \BibitemOpen
  \bibfield  {author} {\bibinfo {author} {\bibfnamefont {A.}~\bibnamefont
  {Li}}\ and\ \bibinfo {author} {\bibfnamefont {S.}~\bibnamefont
  {Krishnamoorthy}},\ }\bibfield  {title} {\bibinfo {title} {{QASMBench}: A
  low-level {QASM} benchmark suite for {NISQ} evaluation and simulation},\
  }\bibfield  {journal} {\bibinfo  {journal} {arXiv preprint arXiv:2005.13018}\
  }\href {https://doi.org/1048550/arXiv.2005.13018} {1048550/arXiv.2005.13018}
  (\bibinfo {year} {2020})\BibitemShut {NoStop}%
\bibitem [{\citenamefont {Quetschlich}\ \emph {et~al.}(2023)\citenamefont
  {Quetschlich}, \citenamefont {Burgholzer},\ and\ \citenamefont
  {Wille}}]{Quetschlich2023-bg}%
  \BibitemOpen
  \bibfield  {author} {\bibinfo {author} {\bibfnamefont {N.}~\bibnamefont
  {Quetschlich}}, \bibinfo {author} {\bibfnamefont {L.}~\bibnamefont
  {Burgholzer}},\ and\ \bibinfo {author} {\bibfnamefont {R.}~\bibnamefont
  {Wille}},\ }\bibfield  {title} {\bibinfo {title} {{MQT} bench: Benchmarking
  software and design automation tools for quantum computing},\ }\href
  {https://doi.org/10.22331/q-2023-07-20-1062} {\bibfield  {journal} {\bibinfo
  {journal} {Quantum}\ }\textbf {\bibinfo {volume} {7}},\ \bibinfo {pages}
  {1062} (\bibinfo {year} {2023})}\BibitemShut {NoStop}%
\bibitem [{\citenamefont {Dong}\ and\ \citenamefont {Lin}(2021)}]{Dong2021-gj}%
  \BibitemOpen
  \bibfield  {author} {\bibinfo {author} {\bibfnamefont {Y.}~\bibnamefont
  {Dong}}\ and\ \bibinfo {author} {\bibfnamefont {L.}~\bibnamefont {Lin}},\
  }\bibfield  {title} {\bibinfo {title} {Random circuit block-encoded matrix
  and a proposal of quantum {LINPACK} benchmark},\ }\href
  {https://doi.org/10.1103/PhysRevA.103.062412} {\bibfield  {journal} {\bibinfo
   {journal} {Phys. Rev. A}\ }\textbf {\bibinfo {volume} {103}},\ \bibinfo
  {pages} {062412} (\bibinfo {year} {2021})}\BibitemShut {NoStop}%
\bibitem [{\citenamefont {Martiel}\ \emph {et~al.}(2021)\citenamefont
  {Martiel}, \citenamefont {Ayral},\ and\ \citenamefont
  {Allouche}}]{Martiel2021-vp}%
  \BibitemOpen
  \bibfield  {author} {\bibinfo {author} {\bibfnamefont {S.}~\bibnamefont
  {Martiel}}, \bibinfo {author} {\bibfnamefont {T.}~\bibnamefont {Ayral}},\
  and\ \bibinfo {author} {\bibfnamefont {C.}~\bibnamefont {Allouche}},\
  }\bibfield  {title} {\bibinfo {title} {Benchmarking quantum coprocessors in
  an {Application-Centric}, {Hardware-Agnostic}, and scalable way},\ }\href
  {https://doi.org/10.1109/TQE.2021.3090207} {\bibfield  {journal} {\bibinfo
  {journal} {IEEE Transactions on Quantum Engineering}\ }\textbf {\bibinfo
  {volume} {2}},\ \bibinfo {pages} {1} (\bibinfo {year} {2021})}\BibitemShut
  {NoStop}%
\bibitem [{\citenamefont {van~der Schoot}\ \emph {et~al.}(2022)\citenamefont
  {van~der Schoot}, \citenamefont {Leermakers}, \citenamefont {Wezeman},
  \citenamefont {Neumann},\ and\ \citenamefont
  {Phillipson}}]{Van_der_Schoot2022-gv}%
  \BibitemOpen
  \bibfield  {author} {\bibinfo {author} {\bibfnamefont {W.}~\bibnamefont
  {van~der Schoot}}, \bibinfo {author} {\bibfnamefont {D.}~\bibnamefont
  {Leermakers}}, \bibinfo {author} {\bibfnamefont {R.}~\bibnamefont {Wezeman}},
  \bibinfo {author} {\bibfnamefont {N.}~\bibnamefont {Neumann}},\ and\ \bibinfo
  {author} {\bibfnamefont {F.}~\bibnamefont {Phillipson}},\ }\bibfield  {title}
  {\bibinfo {title} {Evaluating the q-score of quantum annealers},\ }\bibfield
  {journal} {\bibinfo  {journal} {arXiv preprint arXiv:2208.07633}\ }\href
  {https://doi.org/1048550/arXiv.2208.07633} {1048550/arXiv.2208.07633}
  (\bibinfo {year} {2022})\BibitemShut {NoStop}%
\bibitem [{\citenamefont {van~der Schoot}\ \emph {et~al.}(2023)\citenamefont
  {van~der Schoot}, \citenamefont {Wezeman}, \citenamefont {Neumann},
  \citenamefont {Phillipson},\ and\ \citenamefont
  {Kooij}}]{Van_der_Schoot2023-vo}%
  \BibitemOpen
  \bibfield  {author} {\bibinfo {author} {\bibfnamefont {W.}~\bibnamefont
  {van~der Schoot}}, \bibinfo {author} {\bibfnamefont {R.}~\bibnamefont
  {Wezeman}}, \bibinfo {author} {\bibfnamefont {N.~M.~P.}\ \bibnamefont
  {Neumann}}, \bibinfo {author} {\bibfnamefont {F.}~\bibnamefont
  {Phillipson}},\ and\ \bibinfo {author} {\bibfnamefont {R.}~\bibnamefont
  {Kooij}},\ }\bibfield  {title} {\bibinfo {title} {Q-score {Max-Clique}: The
  first quantum metric evaluation on multiple computational paradigms},\
  }\bibfield  {journal} {\bibinfo  {journal} {arXiv preprint arXiv:2302.00639}\
  }\href {https://doi.org/1048550/arXiv.2302.00639} {1048550/arXiv.2302.00639}
  (\bibinfo {year} {2023})\BibitemShut {NoStop}%
\bibitem [{\citenamefont {Cornelissen}\ \emph {et~al.}(2021)\citenamefont
  {Cornelissen}, \citenamefont {Bausch},\ and\ \citenamefont
  {Gily{\'e}n}}]{Cornelissen2021-yt}%
  \BibitemOpen
  \bibfield  {author} {\bibinfo {author} {\bibfnamefont {A.}~\bibnamefont
  {Cornelissen}}, \bibinfo {author} {\bibfnamefont {J.}~\bibnamefont
  {Bausch}},\ and\ \bibinfo {author} {\bibfnamefont {A.}~\bibnamefont
  {Gily{\'e}n}},\ }\bibfield  {title} {\bibinfo {title} {Scalable benchmarks
  for {Gate-Based} quantum computers},\ }\bibfield  {journal} {\bibinfo
  {journal} {arXiv preprint arXiv:2104.10698}\ }\href
  {https://doi.org/1048550/arXiv.2104.10698} {1048550/arXiv.2104.10698}
  (\bibinfo {year} {2021})\BibitemShut {NoStop}%
\bibitem [{\citenamefont {Georgopoulos}\ \emph {et~al.}(2021)\citenamefont
  {Georgopoulos}, \citenamefont {Emary},\ and\ \citenamefont
  {Zuliani}}]{Georgopoulos2021-hh}%
  \BibitemOpen
  \bibfield  {author} {\bibinfo {author} {\bibfnamefont {K.}~\bibnamefont
  {Georgopoulos}}, \bibinfo {author} {\bibfnamefont {C.}~\bibnamefont
  {Emary}},\ and\ \bibinfo {author} {\bibfnamefont {P.}~\bibnamefont
  {Zuliani}},\ }\bibfield  {title} {\bibinfo {title} {Quantum computer
  benchmarking via quantum algorithms},\ }\bibfield  {journal} {\bibinfo
  {journal} {arXiv preprint arXiv:2112.09457}\ }\href
  {https://doi.org/1048550/arXiv.2112.09457} {1048550/arXiv.2112.09457}
  (\bibinfo {year} {2021})\BibitemShut {NoStop}%
\bibitem [{\citenamefont {Dong}\ \emph {et~al.}(2022)\citenamefont {Dong},
  \citenamefont {Whaley},\ and\ \citenamefont {Lin}}]{Dong2022-ga}%
  \BibitemOpen
  \bibfield  {author} {\bibinfo {author} {\bibfnamefont {Y.}~\bibnamefont
  {Dong}}, \bibinfo {author} {\bibfnamefont {K.~B.}\ \bibnamefont {Whaley}},\
  and\ \bibinfo {author} {\bibfnamefont {L.}~\bibnamefont {Lin}},\ }\bibfield
  {title} {\bibinfo {title} {A quantum hamiltonian simulation benchmark},\
  }\href {https://doi.org/10.1038/s41534-022-00636-x} {\bibfield  {journal}
  {\bibinfo  {journal} {npj Quantum Information}\ }\textbf {\bibinfo {volume}
  {8}},\ \bibinfo {pages} {1} (\bibinfo {year} {2022})}\BibitemShut {NoStop}%
\bibitem [{\citenamefont {Chatterjee}\ \emph {et~al.}(2025)\citenamefont
  {Chatterjee}, \citenamefont {Rappaport}, \citenamefont {Giri}, \citenamefont
  {Johri}, \citenamefont {Proctor}, \citenamefont {Neira}, \citenamefont
  {Sathe},\ and\ \citenamefont {Lubinski}}]{Chatterjee2025-lp}%
  \BibitemOpen
  \bibfield  {author} {\bibinfo {author} {\bibfnamefont {A.}~\bibnamefont
  {Chatterjee}}, \bibinfo {author} {\bibfnamefont {S.}~\bibnamefont
  {Rappaport}}, \bibinfo {author} {\bibfnamefont {A.}~\bibnamefont {Giri}},
  \bibinfo {author} {\bibfnamefont {S.}~\bibnamefont {Johri}}, \bibinfo
  {author} {\bibfnamefont {T.}~\bibnamefont {Proctor}}, \bibinfo {author}
  {\bibfnamefont {D.~E.~B.}\ \bibnamefont {Neira}}, \bibinfo {author}
  {\bibfnamefont {P.}~\bibnamefont {Sathe}},\ and\ \bibinfo {author}
  {\bibfnamefont {T.}~\bibnamefont {Lubinski}},\ }\bibfield  {title} {\bibinfo
  {title} {A comprehensive cross-model framework for benchmarking the
  performance of quantum hamiltonian simulations},\ }\href
  {https://doi.org/10.1109/tqe.2025.3558090} {\bibfield  {journal} {\bibinfo
  {journal} {IEEE Trans. Quantum Eng.}\ }\textbf {\bibinfo {volume} {6}},\
  \bibinfo {pages} {1} (\bibinfo {year} {2025})}\BibitemShut {NoStop}%
\bibitem [{\citenamefont {Gidney}(2025)}]{gidney2025factor}%
  \BibitemOpen
  \bibfield  {author} {\bibinfo {author} {\bibfnamefont {C.}~\bibnamefont
  {Gidney}},\ }\bibfield  {title} {\bibinfo {title} {How to factor 2048 bit rsa
  integers with less than a million noisy qubits},\ }\bibfield  {journal}
  {\bibinfo  {journal} {arXiv preprint arXiv:2505.15917}\ }\href
  {https://doi.org/10.48550/arXiv.2505.15917} {10.48550/arXiv.2505.15917}
  (\bibinfo {year} {2025})\BibitemShut {NoStop}%
\bibitem [{\citenamefont {Low}\ \emph {et~al.}(2025)\citenamefont {Low},
  \citenamefont {King}, \citenamefont {Berry}, \citenamefont {Han},
  \citenamefont {DePrince~III}, \citenamefont {White}, \citenamefont {Babbush},
  \citenamefont {Somma},\ and\ \citenamefont {Rubin}}]{low2025fast}%
  \BibitemOpen
  \bibfield  {author} {\bibinfo {author} {\bibfnamefont {G.~H.}\ \bibnamefont
  {Low}}, \bibinfo {author} {\bibfnamefont {R.}~\bibnamefont {King}}, \bibinfo
  {author} {\bibfnamefont {D.~W.}\ \bibnamefont {Berry}}, \bibinfo {author}
  {\bibfnamefont {Q.}~\bibnamefont {Han}}, \bibinfo {author} {\bibfnamefont
  {A.~E.}\ \bibnamefont {DePrince~III}}, \bibinfo {author} {\bibfnamefont
  {A.}~\bibnamefont {White}}, \bibinfo {author} {\bibfnamefont
  {R.}~\bibnamefont {Babbush}}, \bibinfo {author} {\bibfnamefont {R.~D.}\
  \bibnamefont {Somma}},\ and\ \bibinfo {author} {\bibfnamefont {N.~C.}\
  \bibnamefont {Rubin}},\ }\bibfield  {title} {\bibinfo {title} {Fast quantum
  simulation of electronic structure by spectrum amplification},\ }\bibfield
  {journal} {\bibinfo  {journal} {arXiv preprint arXiv:2502.15882}\ }\href
  {https://doi.org/10.48550/arXiv.2502.15882} {10.48550/arXiv.2502.15882}
  (\bibinfo {year} {2025})\BibitemShut {NoStop}%
\bibitem [{\citenamefont {Rubin}\ \emph {et~al.}(2024)\citenamefont {Rubin},
  \citenamefont {Berry}, \citenamefont {Kononov}, \citenamefont {Malone},
  \citenamefont {Khattar}, \citenamefont {White}, \citenamefont {Lee},
  \citenamefont {Neven}, \citenamefont {Babbush},\ and\ \citenamefont
  {Baczewski}}]{Rubin2024-gc}%
  \BibitemOpen
  \bibfield  {author} {\bibinfo {author} {\bibfnamefont {N.~C.}\ \bibnamefont
  {Rubin}}, \bibinfo {author} {\bibfnamefont {D.~W.}\ \bibnamefont {Berry}},
  \bibinfo {author} {\bibfnamefont {A.}~\bibnamefont {Kononov}}, \bibinfo
  {author} {\bibfnamefont {F.~D.}\ \bibnamefont {Malone}}, \bibinfo {author}
  {\bibfnamefont {T.}~\bibnamefont {Khattar}}, \bibinfo {author} {\bibfnamefont
  {A.}~\bibnamefont {White}}, \bibinfo {author} {\bibfnamefont
  {J.}~\bibnamefont {Lee}}, \bibinfo {author} {\bibfnamefont {H.}~\bibnamefont
  {Neven}}, \bibinfo {author} {\bibfnamefont {R.}~\bibnamefont {Babbush}},\
  and\ \bibinfo {author} {\bibfnamefont {A.~D.}\ \bibnamefont {Baczewski}},\
  }\bibfield  {title} {\bibinfo {title} {Quantum computation of stopping power
  for inertial fusion target design},\ }\href
  {https://doi.org/10.1073/pnas.2317772121} {\bibfield  {journal} {\bibinfo
  {journal} {Proc. Natl. Acad. Sci. U. S. A.}\ }\textbf {\bibinfo {volume}
  {121}},\ \bibinfo {pages} {e2317772121} (\bibinfo {year} {2024})}\BibitemShut
  {NoStop}%
\bibitem [{\citenamefont {Emerson}\ \emph {et~al.}(2005)\citenamefont
  {Emerson}, \citenamefont {Alicki},\ and\ \citenamefont
  {{\.Z}yczkowski}}]{Emerson2005-fd}%
  \BibitemOpen
  \bibfield  {author} {\bibinfo {author} {\bibfnamefont {J.}~\bibnamefont
  {Emerson}}, \bibinfo {author} {\bibfnamefont {R.}~\bibnamefont {Alicki}},\
  and\ \bibinfo {author} {\bibfnamefont {K.}~\bibnamefont {{\.Z}yczkowski}},\
  }\bibfield  {title} {\bibinfo {title} {Scalable noise estimation with random
  unitary operators},\ }\href {https://doi.org/10.1088/1464-4266/7/10/021}
  {\bibfield  {journal} {\bibinfo  {journal} {J. Opt. B Quantum Semiclassical
  Opt.}\ }\textbf {\bibinfo {volume} {7}},\ \bibinfo {pages} {S347} (\bibinfo
  {year} {2005})}\BibitemShut {NoStop}%
\bibitem [{\citenamefont {Emerson}\ \emph {et~al.}(2007)\citenamefont
  {Emerson}, \citenamefont {Silva}, \citenamefont {Moussa}, \citenamefont
  {Ryan}, \citenamefont {Laforest}, \citenamefont {Baugh}, \citenamefont
  {Cory},\ and\ \citenamefont {Laflamme}}]{Emerson2007-am}%
  \BibitemOpen
  \bibfield  {author} {\bibinfo {author} {\bibfnamefont {J.}~\bibnamefont
  {Emerson}}, \bibinfo {author} {\bibfnamefont {M.}~\bibnamefont {Silva}},
  \bibinfo {author} {\bibfnamefont {O.}~\bibnamefont {Moussa}}, \bibinfo
  {author} {\bibfnamefont {C.}~\bibnamefont {Ryan}}, \bibinfo {author}
  {\bibfnamefont {M.}~\bibnamefont {Laforest}}, \bibinfo {author}
  {\bibfnamefont {J.}~\bibnamefont {Baugh}}, \bibinfo {author} {\bibfnamefont
  {D.~G.}\ \bibnamefont {Cory}},\ and\ \bibinfo {author} {\bibfnamefont
  {R.}~\bibnamefont {Laflamme}},\ }\bibfield  {title} {\bibinfo {title}
  {Symmetrized characterization of noisy quantum processes},\ }\href
  {https://doi.org/10.1126/science.1145699} {\bibfield  {journal} {\bibinfo
  {journal} {Science}\ }\textbf {\bibinfo {volume} {317}},\ \bibinfo {pages}
  {1893} (\bibinfo {year} {2007})}\BibitemShut {NoStop}%
\bibitem [{\citenamefont {Knill}\ \emph {et~al.}(2008)\citenamefont {Knill},
  \citenamefont {Leibfried}, \citenamefont {Reichle}, \citenamefont {Britton},
  \citenamefont {Blakestad}, \citenamefont {Jost}, \citenamefont {Langer},
  \citenamefont {Ozeri}, \citenamefont {Seidelin},\ and\ \citenamefont
  {Wineland}}]{Knill2008-jf}%
  \BibitemOpen
  \bibfield  {author} {\bibinfo {author} {\bibfnamefont {E.}~\bibnamefont
  {Knill}}, \bibinfo {author} {\bibfnamefont {D.}~\bibnamefont {Leibfried}},
  \bibinfo {author} {\bibfnamefont {R.}~\bibnamefont {Reichle}}, \bibinfo
  {author} {\bibfnamefont {J.}~\bibnamefont {Britton}}, \bibinfo {author}
  {\bibfnamefont {R.~B.}\ \bibnamefont {Blakestad}}, \bibinfo {author}
  {\bibfnamefont {J.~D.}\ \bibnamefont {Jost}}, \bibinfo {author}
  {\bibfnamefont {C.}~\bibnamefont {Langer}}, \bibinfo {author} {\bibfnamefont
  {R.}~\bibnamefont {Ozeri}}, \bibinfo {author} {\bibfnamefont
  {S.}~\bibnamefont {Seidelin}},\ and\ \bibinfo {author} {\bibfnamefont
  {D.~J.}\ \bibnamefont {Wineland}},\ }\bibfield  {title} {\bibinfo {title}
  {Randomized benchmarking of quantum gates},\ }\href
  {https://doi.org/10.1103/PhysRevA.77.012307} {\bibfield  {journal} {\bibinfo
  {journal} {Phys. Rev. A}\ }\textbf {\bibinfo {volume} {77}},\ \bibinfo
  {pages} {012307} (\bibinfo {year} {2008})}\BibitemShut {NoStop}%
\bibitem [{\citenamefont {Magesan}\ \emph {et~al.}(2011)\citenamefont
  {Magesan}, \citenamefont {Gambetta},\ and\ \citenamefont
  {Emerson}}]{Magesan2011-hc}%
  \BibitemOpen
  \bibfield  {author} {\bibinfo {author} {\bibfnamefont {E.}~\bibnamefont
  {Magesan}}, \bibinfo {author} {\bibfnamefont {J.~M.}\ \bibnamefont
  {Gambetta}},\ and\ \bibinfo {author} {\bibfnamefont {J.}~\bibnamefont
  {Emerson}},\ }\bibfield  {title} {\bibinfo {title} {Scalable and robust
  randomized benchmarking of quantum processes},\ }\href
  {https://doi.org/10.1103/PhysRevLett.106.180504} {\bibfield  {journal}
  {\bibinfo  {journal} {Phys. Rev. Lett.}\ }\textbf {\bibinfo {volume} {106}},\
  \bibinfo {pages} {180504} (\bibinfo {year} {2011})}\BibitemShut {NoStop}%
\bibitem [{\citenamefont {Proctor}\ \emph {et~al.}(2019)\citenamefont
  {Proctor}, \citenamefont {Carignan-Dugas}, \citenamefont {Rudinger},
  \citenamefont {Nielsen}, \citenamefont {Blume-Kohout},\ and\ \citenamefont
  {Young}}]{Proctor2019-gf}%
  \BibitemOpen
  \bibfield  {author} {\bibinfo {author} {\bibfnamefont {T.~J.}\ \bibnamefont
  {Proctor}}, \bibinfo {author} {\bibfnamefont {A.}~\bibnamefont
  {Carignan-Dugas}}, \bibinfo {author} {\bibfnamefont {K.}~\bibnamefont
  {Rudinger}}, \bibinfo {author} {\bibfnamefont {E.}~\bibnamefont {Nielsen}},
  \bibinfo {author} {\bibfnamefont {R.}~\bibnamefont {Blume-Kohout}},\ and\
  \bibinfo {author} {\bibfnamefont {K.}~\bibnamefont {Young}},\ }\bibfield
  {title} {\bibinfo {title} {Direct randomized benchmarking for multiqubit
  devices},\ }\href {https://doi.org/10.1103/PhysRevLett.123.030503} {\bibfield
   {journal} {\bibinfo  {journal} {Phys. Rev. Lett.}\ }\textbf {\bibinfo
  {volume} {123}},\ \bibinfo {pages} {030503} (\bibinfo {year}
  {2019})}\BibitemShut {NoStop}%
\bibitem [{\citenamefont {Sarovar}\ \emph {et~al.}(2020)\citenamefont
  {Sarovar}, \citenamefont {Proctor}, \citenamefont {Rudinger}, \citenamefont
  {Young}, \citenamefont {Nielsen},\ and\ \citenamefont
  {Blume-Kohout}}]{Sarovar2020-pz}%
  \BibitemOpen
  \bibfield  {author} {\bibinfo {author} {\bibfnamefont {M.}~\bibnamefont
  {Sarovar}}, \bibinfo {author} {\bibfnamefont {T.}~\bibnamefont {Proctor}},
  \bibinfo {author} {\bibfnamefont {K.}~\bibnamefont {Rudinger}}, \bibinfo
  {author} {\bibfnamefont {K.}~\bibnamefont {Young}}, \bibinfo {author}
  {\bibfnamefont {E.}~\bibnamefont {Nielsen}},\ and\ \bibinfo {author}
  {\bibfnamefont {R.}~\bibnamefont {Blume-Kohout}},\ }\bibfield  {title}
  {\bibinfo {title} {Detecting crosstalk errors in quantum information
  processors},\ }\href {https://doi.org/10.22331/q-2020-09-11-321} {\bibfield
  {journal} {\bibinfo  {journal} {Quantum}\ }\textbf {\bibinfo {volume} {4}},\
  \bibinfo {pages} {321} (\bibinfo {year} {2020})}\BibitemShut {NoStop}%
\bibitem [{\citenamefont {Gambetta}\ \emph {et~al.}(2012)\citenamefont
  {Gambetta}, \citenamefont {C{\'o}rcoles}, \citenamefont {Merkel},
  \citenamefont {Johnson}, \citenamefont {Smolin}, \citenamefont {Chow},
  \citenamefont {Ryan}, \citenamefont {Rigetti}, \citenamefont {Poletto},
  \citenamefont {Ohki}, \citenamefont {Ketchen},\ and\ \citenamefont
  {Steffen}}]{Gambetta2012-zd}%
  \BibitemOpen
  \bibfield  {author} {\bibinfo {author} {\bibfnamefont {J.~M.}\ \bibnamefont
  {Gambetta}}, \bibinfo {author} {\bibfnamefont {A.~D.}\ \bibnamefont
  {C{\'o}rcoles}}, \bibinfo {author} {\bibfnamefont {S.~T.}\ \bibnamefont
  {Merkel}}, \bibinfo {author} {\bibfnamefont {B.~R.}\ \bibnamefont {Johnson}},
  \bibinfo {author} {\bibfnamefont {J.~A.}\ \bibnamefont {Smolin}}, \bibinfo
  {author} {\bibfnamefont {J.~M.}\ \bibnamefont {Chow}}, \bibinfo {author}
  {\bibfnamefont {C.~A.}\ \bibnamefont {Ryan}}, \bibinfo {author}
  {\bibfnamefont {C.}~\bibnamefont {Rigetti}}, \bibinfo {author} {\bibfnamefont
  {S.}~\bibnamefont {Poletto}}, \bibinfo {author} {\bibfnamefont {T.~A.}\
  \bibnamefont {Ohki}}, \bibinfo {author} {\bibfnamefont {M.~B.}\ \bibnamefont
  {Ketchen}},\ and\ \bibinfo {author} {\bibfnamefont {M.}~\bibnamefont
  {Steffen}},\ }\bibfield  {title} {\bibinfo {title} {Characterization of
  addressability by simultaneous randomized benchmarking},\ }\href
  {https://doi.org/10.1103/PhysRevLett.109.240504} {\bibfield  {journal}
  {\bibinfo  {journal} {Phys. Rev. Lett.}\ }\textbf {\bibinfo {volume} {109}},\
  \bibinfo {pages} {240504} (\bibinfo {year} {2012})}\BibitemShut {NoStop}%
\bibitem [{\citenamefont {Proctor}\ \emph
  {et~al.}(2022{\natexlab{a}})\citenamefont {Proctor}, \citenamefont {Seritan},
  \citenamefont {Rudinger}, \citenamefont {Nielsen}, \citenamefont
  {Blume-Kohout},\ and\ \citenamefont {Young}}]{Proctor2022-yl}%
  \BibitemOpen
  \bibfield  {author} {\bibinfo {author} {\bibfnamefont {T.}~\bibnamefont
  {Proctor}}, \bibinfo {author} {\bibfnamefont {S.}~\bibnamefont {Seritan}},
  \bibinfo {author} {\bibfnamefont {K.}~\bibnamefont {Rudinger}}, \bibinfo
  {author} {\bibfnamefont {E.}~\bibnamefont {Nielsen}}, \bibinfo {author}
  {\bibfnamefont {R.}~\bibnamefont {Blume-Kohout}},\ and\ \bibinfo {author}
  {\bibfnamefont {K.}~\bibnamefont {Young}},\ }\bibfield  {title} {\bibinfo
  {title} {Scalable randomized benchmarking of quantum computers using mirror
  circuits},\ }\href {https://doi.org/10.1103/PhysRevLett.129.150502}
  {\bibfield  {journal} {\bibinfo  {journal} {Phys. Rev. Lett.}\ }\textbf
  {\bibinfo {volume} {129}},\ \bibinfo {pages} {150502} (\bibinfo {year}
  {2022}{\natexlab{a}})}\BibitemShut {NoStop}%
\bibitem [{\citenamefont {Harper}\ and\ \citenamefont
  {Flammia}(2023)}]{Harper2023-pv}%
  \BibitemOpen
  \bibfield  {author} {\bibinfo {author} {\bibfnamefont {R.}~\bibnamefont
  {Harper}}\ and\ \bibinfo {author} {\bibfnamefont {S.~T.}\ \bibnamefont
  {Flammia}},\ }\bibfield  {title} {\bibinfo {title} {Learning correlated noise
  in a 39-qubit quantum processor},\ }\href
  {https://doi.org/10.1103/prxquantum.4.040311} {\bibfield  {journal} {\bibinfo
   {journal} {PRX quantum}\ }\textbf {\bibinfo {volume} {4}},\ \bibinfo {pages}
  {040311} (\bibinfo {year} {2023})}\BibitemShut {NoStop}%
\bibitem [{\citenamefont {Hines}\ and\ \citenamefont
  {Proctor}(2023)}]{Hines2023-be}%
  \BibitemOpen
  \bibfield  {author} {\bibinfo {author} {\bibfnamefont {J.}~\bibnamefont
  {Hines}}\ and\ \bibinfo {author} {\bibfnamefont {T.}~\bibnamefont
  {Proctor}},\ }\bibfield  {title} {\bibinfo {title} {Scalable {Full-Stack}
  benchmarks for quantum computers},\ }\bibfield  {journal} {\bibinfo
  {journal} {arXiv preprint arXiv:2312.14107}\ }\href
  {https://doi.org/1048550/arXiv.2312.14107} {1048550/arXiv.2312.14107}
  (\bibinfo {year} {2023})\BibitemShut {NoStop}%
\bibitem [{\citenamefont {Smolin}\ \emph {et~al.}(2013)\citenamefont {Smolin},
  \citenamefont {Smith},\ and\ \citenamefont {Vargo}}]{Smolin2013-vc}%
  \BibitemOpen
  \bibfield  {author} {\bibinfo {author} {\bibfnamefont {J.~A.}\ \bibnamefont
  {Smolin}}, \bibinfo {author} {\bibfnamefont {G.}~\bibnamefont {Smith}},\ and\
  \bibinfo {author} {\bibfnamefont {A.}~\bibnamefont {Vargo}},\ }\bibfield
  {title} {\bibinfo {title} {Oversimplifying quantum factoring},\ }\href
  {https://doi.org/10.1038/nature12290} {\bibfield  {journal} {\bibinfo
  {journal} {Nature}\ }\textbf {\bibinfo {volume} {499}},\ \bibinfo {pages}
  {163} (\bibinfo {year} {2013})}\BibitemShut {NoStop}%
\bibitem [{\citenamefont {Low}\ and\ \citenamefont
  {Chuang}(2017)}]{low2017hamiltonian}%
  \BibitemOpen
  \bibfield  {author} {\bibinfo {author} {\bibfnamefont {G.~H.}\ \bibnamefont
  {Low}}\ and\ \bibinfo {author} {\bibfnamefont {I.~L.}\ \bibnamefont
  {Chuang}},\ }\bibfield  {title} {\bibinfo {title} {Hamiltonian simulation by
  uniform spectral amplification},\ }\bibfield  {journal} {\bibinfo  {journal}
  {arXiv preprint arXiv:1707.05391}\ }\href
  {https://doi.org/10.48550/arXiv.1707.05391} {10.48550/arXiv.1707.05391}
  (\bibinfo {year} {2017})\BibitemShut {NoStop}%
\bibitem [{\citenamefont {Chakraborty}\ \emph {et~al.}(2018)\citenamefont
  {Chakraborty}, \citenamefont {Gily{\'e}n},\ and\ \citenamefont
  {Jeffery}}]{chakraborty2018power}%
  \BibitemOpen
  \bibfield  {author} {\bibinfo {author} {\bibfnamefont {S.}~\bibnamefont
  {Chakraborty}}, \bibinfo {author} {\bibfnamefont {A.}~\bibnamefont
  {Gily{\'e}n}},\ and\ \bibinfo {author} {\bibfnamefont {S.}~\bibnamefont
  {Jeffery}},\ }\bibfield  {title} {\bibinfo {title} {The power of
  block-encoded matrix powers: improved regression techniques via faster
  hamiltonian simulation},\ }\bibfield  {journal} {\bibinfo  {journal} {arXiv
  preprint arXiv:1804.01973}\ }\href
  {https://doi.org/10.48550/arXiv.1804.01973} {10.48550/arXiv.1804.01973}
  (\bibinfo {year} {2018})\BibitemShut {NoStop}%
\bibitem [{\citenamefont {Gily{\'e}n}\ \emph {et~al.}(2019)\citenamefont
  {Gily{\'e}n}, \citenamefont {Su}, \citenamefont {Low},\ and\ \citenamefont
  {Wiebe}}]{gilyen2019quantum}%
  \BibitemOpen
  \bibfield  {author} {\bibinfo {author} {\bibfnamefont {A.}~\bibnamefont
  {Gily{\'e}n}}, \bibinfo {author} {\bibfnamefont {Y.}~\bibnamefont {Su}},
  \bibinfo {author} {\bibfnamefont {G.~H.}\ \bibnamefont {Low}},\ and\ \bibinfo
  {author} {\bibfnamefont {N.}~\bibnamefont {Wiebe}},\ }\bibfield  {title}
  {\bibinfo {title} {Quantum singular value transformation and beyond:
  exponential improvements for quantum matrix arithmetics},\ }in\ \href
  {https://doi.org/10.1145/3313276.3316366} {\emph {\bibinfo {booktitle}
  {Proceedings of the 51st annual ACM SIGACT symposium on theory of
  computing}}}\ (\bibinfo {year} {2019})\ pp.\ \bibinfo {pages}
  {193--204}\BibitemShut {NoStop}%
\bibitem [{Note1()}]{Note1}%
  \BibitemOpen
  \bibinfo {note} {That is the qubits in $q_w$ correspond to a connected
  sub-graph of $\protect \mathcal {Q}$'s connectivity graph.}\BibitemShut
  {Stop}%
\bibitem [{Note2()}]{Note2}%
  \BibitemOpen
  \bibinfo {note} {We also relabel the qubits in our snippet, according to the
  equivalence mapping from our subset of $\protect \mathcal {Q}$ to $\protect
  \mathcal {Q}'$}\BibitemShut {NoStop}%
\bibitem [{\citenamefont {Proctor}\ \emph
  {et~al.}(2022{\natexlab{b}})\citenamefont {Proctor}, \citenamefont {Seritan},
  \citenamefont {Nielsen}, \citenamefont {Rudinger}, \citenamefont {Young},
  \citenamefont {Blume-Kohout},\ and\ \citenamefont
  {Sarovar}}]{Proctor2022-zs}%
  \BibitemOpen
  \bibfield  {author} {\bibinfo {author} {\bibfnamefont {T.}~\bibnamefont
  {Proctor}}, \bibinfo {author} {\bibfnamefont {S.}~\bibnamefont {Seritan}},
  \bibinfo {author} {\bibfnamefont {E.}~\bibnamefont {Nielsen}}, \bibinfo
  {author} {\bibfnamefont {K.}~\bibnamefont {Rudinger}}, \bibinfo {author}
  {\bibfnamefont {K.}~\bibnamefont {Young}}, \bibinfo {author} {\bibfnamefont
  {R.}~\bibnamefont {Blume-Kohout}},\ and\ \bibinfo {author} {\bibfnamefont
  {M.}~\bibnamefont {Sarovar}},\ }\bibfield  {title} {\bibinfo {title}
  {Establishing trust in quantum computations},\ }\bibfield  {journal}
  {\bibinfo  {journal} {arXiv preprint arXiv:2204.07568}\ }\href
  {https://doi.org/1048550/arXiv.2204.07568} {1048550/arXiv.2204.07568}
  (\bibinfo {year} {2022}{\natexlab{b}})\BibitemShut {NoStop}%
\bibitem [{\citenamefont {Seth}\ \emph {et~al.}(2025)\citenamefont {Seth},
  \citenamefont {Timothy}, \citenamefont {Samuele}, \citenamefont {Jordan},
  \citenamefont {Samantha}, \citenamefont {Govia},\ and\ \citenamefont
  {David}}]{Seth2025-zz}%
  \BibitemOpen
  \bibfield  {author} {\bibinfo {author} {\bibfnamefont {M.}~\bibnamefont
  {Seth}}, \bibinfo {author} {\bibfnamefont {P.}~\bibnamefont {Timothy}},
  \bibinfo {author} {\bibfnamefont {F.}~\bibnamefont {Samuele}}, \bibinfo
  {author} {\bibfnamefont {H.}~\bibnamefont {Jordan}}, \bibinfo {author}
  {\bibfnamefont {B.}~\bibnamefont {Samantha}}, \bibinfo {author}
  {\bibfnamefont {L.~C.~G.}\ \bibnamefont {Govia}},\ and\ \bibinfo {author}
  {\bibfnamefont {M.}~\bibnamefont {David}},\ }\bibfield  {title} {\bibinfo
  {title} {When clifford benchmarks are sufficient; estimating application
  performance with scalable proxy circuits},\ }\href
  {http://arxiv.org/abs/2503.05943} {\bibfield  {journal} {\bibinfo  {journal}
  {arXiv [quant-ph]}\ } (\bibinfo {year} {2025})},\ \Eprint
  {https://arxiv.org/abs/2503.05943} {arXiv:2503.05943 [quant-ph]} \BibitemShut
  {NoStop}%
\bibitem [{Note3()}]{Note3}%
  \BibitemOpen
  \bibinfo {note} {There are two sources of approximation. First, there is an
  $\protect \mathcal {O}(1/4^w)$ correction factor that is related to the
  difference between process fidelity and process polarization \cite
  {Hashim2024-om} (which, because we are typically interested in $w \gg 1$, we
  ignore for simplicity). Second, stochastic errors can cancel out within a
  circuit, and the exact rate of cancellation varies from circuit to circuit
  and depends on the exact biases in the errors. The impact of this effect on
  $F_{w,d}$ is at most $\protect \mathcal {O}(\epsilon ^2 w^2 d^2)$ but, in
  practice, it causes a negligible correction to Eq.~\protect \eqref
  {eq:f_approx}.}\BibitemShut {Stop}%
\bibitem [{\citenamefont {Ferracin}\ \emph {et~al.}(2021)\citenamefont
  {Ferracin}, \citenamefont {Merkel}, \citenamefont {McKay},\ and\
  \citenamefont {Datta}}]{Ferracin2021-vh}%
  \BibitemOpen
  \bibfield  {author} {\bibinfo {author} {\bibfnamefont {S.}~\bibnamefont
  {Ferracin}}, \bibinfo {author} {\bibfnamefont {S.~T.}\ \bibnamefont
  {Merkel}}, \bibinfo {author} {\bibfnamefont {D.}~\bibnamefont {McKay}},\ and\
  \bibinfo {author} {\bibfnamefont {A.}~\bibnamefont {Datta}},\ }\bibfield
  {title} {\bibinfo {title} {Experimental accreditation of outputs of noisy
  quantum computers},\ }\href {https://doi.org/10.1103/PhysRevA.104.042603}
  {\bibfield  {journal} {\bibinfo  {journal} {Phys. Rev. A}\ }\textbf {\bibinfo
  {volume} {104}},\ \bibinfo {pages} {042603} (\bibinfo {year}
  {2021})}\BibitemShut {NoStop}%
\bibitem [{\citenamefont {Cao}\ \emph {et~al.}(2019)\citenamefont {Cao},
  \citenamefont {Romero}, \citenamefont {Olson}, \citenamefont {Degroote},
  \citenamefont {Johnson}, \citenamefont {Kieferov{\'a}}, \citenamefont
  {Kivlichan}, \citenamefont {Menke}, \citenamefont {Peropadre}, \citenamefont
  {Sawaya} \emph {et~al.}}]{cao2019quantum}%
  \BibitemOpen
  \bibfield  {author} {\bibinfo {author} {\bibfnamefont {Y.}~\bibnamefont
  {Cao}}, \bibinfo {author} {\bibfnamefont {J.}~\bibnamefont {Romero}},
  \bibinfo {author} {\bibfnamefont {J.~P.}\ \bibnamefont {Olson}}, \bibinfo
  {author} {\bibfnamefont {M.}~\bibnamefont {Degroote}}, \bibinfo {author}
  {\bibfnamefont {P.~D.}\ \bibnamefont {Johnson}}, \bibinfo {author}
  {\bibfnamefont {M.}~\bibnamefont {Kieferov{\'a}}}, \bibinfo {author}
  {\bibfnamefont {I.~D.}\ \bibnamefont {Kivlichan}}, \bibinfo {author}
  {\bibfnamefont {T.}~\bibnamefont {Menke}}, \bibinfo {author} {\bibfnamefont
  {B.}~\bibnamefont {Peropadre}}, \bibinfo {author} {\bibfnamefont {N.~P.}\
  \bibnamefont {Sawaya}}, \emph {et~al.},\ }\bibfield  {title} {\bibinfo
  {title} {Quantum chemistry in the age of quantum computing},\ }\href
  {https://doi.org/10.1021/acs.chemrev.8b00803} {\bibfield  {journal} {\bibinfo
   {journal} {Chemical reviews}\ }\textbf {\bibinfo {volume} {119}},\ \bibinfo
  {pages} {10856} (\bibinfo {year} {2019})}\BibitemShut {NoStop}%
\bibitem [{\citenamefont {McArdle}\ \emph {et~al.}(2020)\citenamefont
  {McArdle}, \citenamefont {Endo}, \citenamefont {Aspuru-Guzik}, \citenamefont
  {Benjamin},\ and\ \citenamefont {Yuan}}]{mcardle2020quantum}%
  \BibitemOpen
  \bibfield  {author} {\bibinfo {author} {\bibfnamefont {S.}~\bibnamefont
  {McArdle}}, \bibinfo {author} {\bibfnamefont {S.}~\bibnamefont {Endo}},
  \bibinfo {author} {\bibfnamefont {A.}~\bibnamefont {Aspuru-Guzik}}, \bibinfo
  {author} {\bibfnamefont {S.~C.}\ \bibnamefont {Benjamin}},\ and\ \bibinfo
  {author} {\bibfnamefont {X.}~\bibnamefont {Yuan}},\ }\bibfield  {title}
  {\bibinfo {title} {Quantum computational chemistry},\ }\href
  {https://doi.org/10.1103/RevModPhys.92.015003} {\bibfield  {journal}
  {\bibinfo  {journal} {Reviews of Modern Physics}\ }\textbf {\bibinfo {volume}
  {92}},\ \bibinfo {pages} {015003} (\bibinfo {year} {2020})}\BibitemShut
  {NoStop}%
\bibitem [{\citenamefont {Aspuru-Guzik}\ \emph {et~al.}(2005)\citenamefont
  {Aspuru-Guzik}, \citenamefont {Dutoi}, \citenamefont {Love},\ and\
  \citenamefont {Head-Gordon}}]{aspuru2005simulated}%
  \BibitemOpen
  \bibfield  {author} {\bibinfo {author} {\bibfnamefont {A.}~\bibnamefont
  {Aspuru-Guzik}}, \bibinfo {author} {\bibfnamefont {A.~D.}\ \bibnamefont
  {Dutoi}}, \bibinfo {author} {\bibfnamefont {P.~J.}\ \bibnamefont {Love}},\
  and\ \bibinfo {author} {\bibfnamefont {M.}~\bibnamefont {Head-Gordon}},\
  }\bibfield  {title} {\bibinfo {title} {Simulated quantum computation of
  molecular energies},\ }\href {https://doi.org/10.1126/science.1113479}
  {\bibfield  {journal} {\bibinfo  {journal} {Science}\ }\textbf {\bibinfo
  {volume} {309}},\ \bibinfo {pages} {1704} (\bibinfo {year}
  {2005})}\BibitemShut {NoStop}%
\bibitem [{\citenamefont {Pathak}\ \emph {et~al.}(2023)\citenamefont {Pathak},
  \citenamefont {Russo}, \citenamefont {Seritan},\ and\ \citenamefont
  {Baczewski}}]{pathak2023quantifying}%
  \BibitemOpen
  \bibfield  {author} {\bibinfo {author} {\bibfnamefont {S.}~\bibnamefont
  {Pathak}}, \bibinfo {author} {\bibfnamefont {A.~E.}\ \bibnamefont {Russo}},
  \bibinfo {author} {\bibfnamefont {S.~K.}\ \bibnamefont {Seritan}},\ and\
  \bibinfo {author} {\bibfnamefont {A.~D.}\ \bibnamefont {Baczewski}},\
  }\bibfield  {title} {\bibinfo {title} {Quantifying {T}-gate-count
  improvements for ground-state-energy estimation with near-optimal state
  preparation},\ }\href {https://doi.org/10.1103/PhysRevA.107.L040601}
  {\bibfield  {journal} {\bibinfo  {journal} {Physical Review A}\ }\textbf
  {\bibinfo {volume} {107}},\ \bibinfo {pages} {L040601} (\bibinfo {year}
  {2023})}\BibitemShut {NoStop}%
\bibitem [{\citenamefont {Berry}\ \emph {et~al.}(2025)\citenamefont {Berry},
  \citenamefont {Tong}, \citenamefont {Khattar}, \citenamefont {White},
  \citenamefont {Kim}, \citenamefont {Low}, \citenamefont {Boixo},
  \citenamefont {Ding}, \citenamefont {Lin}, \citenamefont {Lee} \emph
  {et~al.}}]{berry2025rapid}%
  \BibitemOpen
  \bibfield  {author} {\bibinfo {author} {\bibfnamefont {D.~W.}\ \bibnamefont
  {Berry}}, \bibinfo {author} {\bibfnamefont {Y.}~\bibnamefont {Tong}},
  \bibinfo {author} {\bibfnamefont {T.}~\bibnamefont {Khattar}}, \bibinfo
  {author} {\bibfnamefont {A.}~\bibnamefont {White}}, \bibinfo {author}
  {\bibfnamefont {T.~I.}\ \bibnamefont {Kim}}, \bibinfo {author} {\bibfnamefont
  {G.~H.}\ \bibnamefont {Low}}, \bibinfo {author} {\bibfnamefont
  {S.}~\bibnamefont {Boixo}}, \bibinfo {author} {\bibfnamefont
  {Z.}~\bibnamefont {Ding}}, \bibinfo {author} {\bibfnamefont {L.}~\bibnamefont
  {Lin}}, \bibinfo {author} {\bibfnamefont {S.}~\bibnamefont {Lee}}, \emph
  {et~al.},\ }\bibfield  {title} {\bibinfo {title} {Rapid initial-state
  preparation for the quantum simulation of strongly correlated molecules},\
  }\href {https://doi.org/10.1103/PRXQuantum.6.020327} {\bibfield  {journal}
  {\bibinfo  {journal} {PRX Quantum}\ }\textbf {\bibinfo {volume} {6}},\
  \bibinfo {pages} {020327} (\bibinfo {year} {2025})}\BibitemShut {NoStop}%
\bibitem [{\citenamefont {Babbush}\ \emph {et~al.}(2018)\citenamefont
  {Babbush}, \citenamefont {Gidney}, \citenamefont {Berry}, \citenamefont
  {Wiebe}, \citenamefont {McClean}, \citenamefont {Paler}, \citenamefont
  {Fowler},\ and\ \citenamefont {Neven}}]{babbush2018encoding}%
  \BibitemOpen
  \bibfield  {author} {\bibinfo {author} {\bibfnamefont {R.}~\bibnamefont
  {Babbush}}, \bibinfo {author} {\bibfnamefont {C.}~\bibnamefont {Gidney}},
  \bibinfo {author} {\bibfnamefont {D.~W.}\ \bibnamefont {Berry}}, \bibinfo
  {author} {\bibfnamefont {N.}~\bibnamefont {Wiebe}}, \bibinfo {author}
  {\bibfnamefont {J.}~\bibnamefont {McClean}}, \bibinfo {author} {\bibfnamefont
  {A.}~\bibnamefont {Paler}}, \bibinfo {author} {\bibfnamefont
  {A.}~\bibnamefont {Fowler}},\ and\ \bibinfo {author} {\bibfnamefont
  {H.}~\bibnamefont {Neven}},\ }\bibfield  {title} {\bibinfo {title} {Encoding
  electronic spectra in quantum circuits with linear {T} complexity},\ }\href
  {https://doi.org/10.1103/PhysRevX.8.041015} {\bibfield  {journal} {\bibinfo
  {journal} {Physical Review X}\ }\textbf {\bibinfo {volume} {8}},\ \bibinfo
  {pages} {041015} (\bibinfo {year} {2018})}\BibitemShut {NoStop}%
\bibitem [{\citenamefont {Russo}\ \emph {et~al.}(2021)\citenamefont {Russo},
  \citenamefont {Rudinger}, \citenamefont {Morrison},\ and\ \citenamefont
  {Baczewski}}]{russo2021evaluating}%
  \BibitemOpen
  \bibfield  {author} {\bibinfo {author} {\bibfnamefont {A.~E.}\ \bibnamefont
  {Russo}}, \bibinfo {author} {\bibfnamefont {K.~M.}\ \bibnamefont {Rudinger}},
  \bibinfo {author} {\bibfnamefont {B.~C.}\ \bibnamefont {Morrison}},\ and\
  \bibinfo {author} {\bibfnamefont {A.~D.}\ \bibnamefont {Baczewski}},\
  }\bibfield  {title} {\bibinfo {title} {Evaluating energy differences on a
  quantum computer with robust phase estimation},\ }\href
  {https://doi.org/10.1103/PhysRevLett.126.210501} {\bibfield  {journal}
  {\bibinfo  {journal} {Physical review letters}\ }\textbf {\bibinfo {volume}
  {126}},\ \bibinfo {pages} {210501} (\bibinfo {year} {2021})}\BibitemShut
  {NoStop}%
\bibitem [{\citenamefont {Martyn}\ \emph {et~al.}(2021)\citenamefont {Martyn},
  \citenamefont {Rossi}, \citenamefont {Tan},\ and\ \citenamefont
  {Chuang}}]{martyn2021grand}%
  \BibitemOpen
  \bibfield  {author} {\bibinfo {author} {\bibfnamefont {J.~M.}\ \bibnamefont
  {Martyn}}, \bibinfo {author} {\bibfnamefont {Z.~M.}\ \bibnamefont {Rossi}},
  \bibinfo {author} {\bibfnamefont {A.~K.}\ \bibnamefont {Tan}},\ and\ \bibinfo
  {author} {\bibfnamefont {I.~L.}\ \bibnamefont {Chuang}},\ }\bibfield  {title}
  {\bibinfo {title} {Grand unification of quantum algorithms},\ }\href
  {https://doi.org/10.1103/PRXQuantum.2.040203} {\bibfield  {journal} {\bibinfo
   {journal} {PRX quantum}\ }\textbf {\bibinfo {volume} {2}},\ \bibinfo {pages}
  {040203} (\bibinfo {year} {2021})}\BibitemShut {NoStop}%
\bibitem [{\citenamefont {Nelson}\ and\ \citenamefont
  {Baczewski}(2024)}]{nelson2024assessment}%
  \BibitemOpen
  \bibfield  {author} {\bibinfo {author} {\bibfnamefont {J.~S.}\ \bibnamefont
  {Nelson}}\ and\ \bibinfo {author} {\bibfnamefont {A.~D.}\ \bibnamefont
  {Baczewski}},\ }\bibfield  {title} {\bibinfo {title} {Assessment of quantum
  phase estimation protocols for early fault-tolerant quantum computers},\
  }\href {https://doi.org/10.1103/PhysRevA.110.042420} {\bibfield  {journal}
  {\bibinfo  {journal} {Physical Review A}\ }\textbf {\bibinfo {volume}
  {110}},\ \bibinfo {pages} {042420} (\bibinfo {year} {2024})}\BibitemShut
  {NoStop}%
\bibitem [{\citenamefont {Low}\ and\ \citenamefont
  {Chuang}(2019)}]{low2019hamiltonian}%
  \BibitemOpen
  \bibfield  {author} {\bibinfo {author} {\bibfnamefont {G.~H.}\ \bibnamefont
  {Low}}\ and\ \bibinfo {author} {\bibfnamefont {I.~L.}\ \bibnamefont
  {Chuang}},\ }\bibfield  {title} {\bibinfo {title} {Hamiltonian simulation by
  qubitization},\ }\href {https://doi.org/10.22331/q-2019-07-12-163} {\bibfield
   {journal} {\bibinfo  {journal} {Quantum}\ }\textbf {\bibinfo {volume} {3}},\
  \bibinfo {pages} {163} (\bibinfo {year} {2019})}\BibitemShut {NoStop}%
\bibitem [{\citenamefont {King}\ \emph {et~al.}(2025)\citenamefont {King},
  \citenamefont {Low}, \citenamefont {Babbush}, \citenamefont {Somma},\ and\
  \citenamefont {Rubin}}]{king2025quantum}%
  \BibitemOpen
  \bibfield  {author} {\bibinfo {author} {\bibfnamefont {R.}~\bibnamefont
  {King}}, \bibinfo {author} {\bibfnamefont {G.~H.}\ \bibnamefont {Low}},
  \bibinfo {author} {\bibfnamefont {R.}~\bibnamefont {Babbush}}, \bibinfo
  {author} {\bibfnamefont {R.~D.}\ \bibnamefont {Somma}},\ and\ \bibinfo
  {author} {\bibfnamefont {N.~C.}\ \bibnamefont {Rubin}},\ }\bibfield  {title}
  {\bibinfo {title} {Quantum simulation with sum-of-squares spectral
  amplification},\ }\bibfield  {journal} {\bibinfo  {journal} {arXiv preprint
  arXiv:2505.01528}\ }\href {https://doi.org/10.48550/arXiv.2505.01528}
  {10.48550/arXiv.2505.01528} (\bibinfo {year} {2025})\BibitemShut {NoStop}%
\bibitem [{\citenamefont {Childs}\ and\ \citenamefont
  {Wiebe}(2012)}]{childs2012hamiltonian}%
  \BibitemOpen
  \bibfield  {author} {\bibinfo {author} {\bibfnamefont {A.~M.}\ \bibnamefont
  {Childs}}\ and\ \bibinfo {author} {\bibfnamefont {N.}~\bibnamefont {Wiebe}},\
  }\bibfield  {title} {\bibinfo {title} {Hamiltonian simulation using linear
  combinations of unitary operations},\ }\href
  {https://doi.org/10.5555/2481569.2481570} {\bibfield  {journal} {\bibinfo
  {journal} {Quantum Information and Computation}\ }\textbf {\bibinfo {volume}
  {12}},\ \bibinfo {pages} {901} (\bibinfo {year} {2012})}\BibitemShut
  {NoStop}%
\bibitem [{\citenamefont {Hempel}\ \emph {et~al.}(2018)\citenamefont {Hempel},
  \citenamefont {Maier}, \citenamefont {Romero}, \citenamefont {McClean},
  \citenamefont {Monz}, \citenamefont {Shen}, \citenamefont {Jurcevic},
  \citenamefont {Lanyon}, \citenamefont {Love}, \citenamefont {Babbush},
  \citenamefont {Aspuru-Guzik}, \citenamefont {Blatt},\ and\ \citenamefont
  {Roos}}]{hempel2018quantum}%
  \BibitemOpen
  \bibfield  {author} {\bibinfo {author} {\bibfnamefont {C.}~\bibnamefont
  {Hempel}}, \bibinfo {author} {\bibfnamefont {C.}~\bibnamefont {Maier}},
  \bibinfo {author} {\bibfnamefont {J.}~\bibnamefont {Romero}}, \bibinfo
  {author} {\bibfnamefont {J.}~\bibnamefont {McClean}}, \bibinfo {author}
  {\bibfnamefont {T.}~\bibnamefont {Monz}}, \bibinfo {author} {\bibfnamefont
  {H.}~\bibnamefont {Shen}}, \bibinfo {author} {\bibfnamefont {P.}~\bibnamefont
  {Jurcevic}}, \bibinfo {author} {\bibfnamefont {B.~P.}\ \bibnamefont
  {Lanyon}}, \bibinfo {author} {\bibfnamefont {P.}~\bibnamefont {Love}},
  \bibinfo {author} {\bibfnamefont {R.}~\bibnamefont {Babbush}}, \bibinfo
  {author} {\bibfnamefont {A.}~\bibnamefont {Aspuru-Guzik}}, \bibinfo {author}
  {\bibfnamefont {R.}~\bibnamefont {Blatt}},\ and\ \bibinfo {author}
  {\bibfnamefont {C.~F.}\ \bibnamefont {Roos}},\ }\bibfield  {title} {\bibinfo
  {title} {{Quantum Chemistry Calculations on a Trapped-Ion Quantum
  Simulator}},\ }\href {https://doi.org/10.1103/physrevx.8.031022} {\bibfield
  {journal} {\bibinfo  {journal} {Phys. Rev. X}\ }\textbf {\bibinfo {volume}
  {8}},\ \bibinfo {pages} {031022} (\bibinfo {year} {2018})},\ \Eprint
  {https://arxiv.org/abs/1803.10238} {1803.10238} \BibitemShut {NoStop}%
\bibitem [{\citenamefont {Maupin}\ \emph {et~al.}(2021)\citenamefont {Maupin},
  \citenamefont {Baczewski}, \citenamefont {Love},\ and\ \citenamefont
  {Landahl}}]{maupin2021variational}%
  \BibitemOpen
  \bibfield  {author} {\bibinfo {author} {\bibfnamefont {O.~G.}\ \bibnamefont
  {Maupin}}, \bibinfo {author} {\bibfnamefont {A.~D.}\ \bibnamefont
  {Baczewski}}, \bibinfo {author} {\bibfnamefont {P.~J.}\ \bibnamefont
  {Love}},\ and\ \bibinfo {author} {\bibfnamefont {A.~J.}\ \bibnamefont
  {Landahl}},\ }\bibfield  {title} {\bibinfo {title} {{Variational Quantum
  Chemistry Programs in JaqalPaq}},\ }\href {https://doi.org/10.3390/e23060657}
  {\bibfield  {journal} {\bibinfo  {journal} {Entropy}\ }\textbf {\bibinfo
  {volume} {23}},\ \bibinfo {pages} {657} (\bibinfo {year} {2021})}\BibitemShut
  {NoStop}%
\bibitem [{\citenamefont {Bravyi}\ and\ \citenamefont
  {Kitaev}(2002)}]{bravyi2002fermionic}%
  \BibitemOpen
  \bibfield  {author} {\bibinfo {author} {\bibfnamefont {S.~B.}\ \bibnamefont
  {Bravyi}}\ and\ \bibinfo {author} {\bibfnamefont {A.~Y.}\ \bibnamefont
  {Kitaev}},\ }\bibfield  {title} {\bibinfo {title} {Fermionic quantum
  computation},\ }\href {https://doi.org/10.1006/aphy.2002.6254} {\bibfield
  {journal} {\bibinfo  {journal} {Annals of Physics}\ }\textbf {\bibinfo
  {volume} {298}},\ \bibinfo {pages} {210} (\bibinfo {year}
  {2002})}\BibitemShut {NoStop}%
\bibitem [{\citenamefont {Jordan}\ and\ \citenamefont
  {Wigner}(1928)}]{jordan1928uber}%
  \BibitemOpen
  \bibfield  {author} {\bibinfo {author} {\bibfnamefont {P.}~\bibnamefont
  {Jordan}}\ and\ \bibinfo {author} {\bibfnamefont {E.}~\bibnamefont
  {Wigner}},\ }\bibfield  {title} {\bibinfo {title} {{Über das Paulische
  Äquivalenzverbot}},\ }\href {https://doi.org/10.1007/bf01331938} {\bibfield
  {journal} {\bibinfo  {journal} {Zeitschrift für Physik}\ }\textbf {\bibinfo
  {volume} {47}},\ \bibinfo {pages} {631} (\bibinfo {year} {1928})}\BibitemShut
  {NoStop}%
\bibitem [{\citenamefont {Bravyi}\ \emph {et~al.}(2017)\citenamefont {Bravyi},
  \citenamefont {Gambetta}, \citenamefont {Mezzacapo},\ and\ \citenamefont
  {Temme}}]{bravyi2017tapering}%
  \BibitemOpen
  \bibfield  {author} {\bibinfo {author} {\bibfnamefont {S.}~\bibnamefont
  {Bravyi}}, \bibinfo {author} {\bibfnamefont {J.~M.}\ \bibnamefont
  {Gambetta}}, \bibinfo {author} {\bibfnamefont {A.}~\bibnamefont
  {Mezzacapo}},\ and\ \bibinfo {author} {\bibfnamefont {K.}~\bibnamefont
  {Temme}},\ }\bibfield  {title} {\bibinfo {title} {Tapering off qubits to
  simulate fermionic hamiltonians},\ }\bibfield  {journal} {\bibinfo  {journal}
  {arXiv preprint arXiv:1701.08213}\ }\href
  {https://doi.org/10.48550/arXiv.1701.08213} {10.48550/arXiv.1701.08213}
  (\bibinfo {year} {2017})\BibitemShut {NoStop}%
\bibitem [{\citenamefont {Kim}\ \emph {et~al.}(2023)\citenamefont {Kim},
  \citenamefont {Eddins}, \citenamefont {Anand}, \citenamefont {Wei},
  \citenamefont {van~den Berg}, \citenamefont {Rosenblatt}, \citenamefont
  {Nayfeh}, \citenamefont {Wu}, \citenamefont {Zaletel}, \citenamefont
  {Temme},\ and\ \citenamefont {Kandala}}]{Kim2023-si}%
  \BibitemOpen
  \bibfield  {author} {\bibinfo {author} {\bibfnamefont {Y.}~\bibnamefont
  {Kim}}, \bibinfo {author} {\bibfnamefont {A.}~\bibnamefont {Eddins}},
  \bibinfo {author} {\bibfnamefont {S.}~\bibnamefont {Anand}}, \bibinfo
  {author} {\bibfnamefont {K.~X.}\ \bibnamefont {Wei}}, \bibinfo {author}
  {\bibfnamefont {E.}~\bibnamefont {van~den Berg}}, \bibinfo {author}
  {\bibfnamefont {S.}~\bibnamefont {Rosenblatt}}, \bibinfo {author}
  {\bibfnamefont {H.}~\bibnamefont {Nayfeh}}, \bibinfo {author} {\bibfnamefont
  {Y.}~\bibnamefont {Wu}}, \bibinfo {author} {\bibfnamefont {M.}~\bibnamefont
  {Zaletel}}, \bibinfo {author} {\bibfnamefont {K.}~\bibnamefont {Temme}},\
  and\ \bibinfo {author} {\bibfnamefont {A.}~\bibnamefont {Kandala}},\
  }\bibfield  {title} {\bibinfo {title} {Evidence for the utility of quantum
  computing before fault tolerance},\ }\href
  {https://doi.org/10.1038/s41586-023-06096-3} {\bibfield  {journal} {\bibinfo
  {journal} {Nature}\ }\textbf {\bibinfo {volume} {618}},\ \bibinfo {pages}
  {500} (\bibinfo {year} {2023})}\BibitemShut {NoStop}%
\bibitem [{Note4()}]{Note4}%
  \BibitemOpen
  \bibinfo {note} {We do not distinguish between idle operations and gates,
  because, in real systems, idle operations are often as noisy as gates and
  cannot be \protect \emph {a priori} assumed to have very low
  error.}\BibitemShut {Stop}%
\bibitem [{\citenamefont {Campbell}\ \emph {et~al.}(2017)\citenamefont
  {Campbell}, \citenamefont {Terhal},\ and\ \citenamefont
  {Vuillot}}]{Campbell2017-tw}%
  \BibitemOpen
  \bibfield  {author} {\bibinfo {author} {\bibfnamefont {E.~T.}\ \bibnamefont
  {Campbell}}, \bibinfo {author} {\bibfnamefont {B.~M.}\ \bibnamefont
  {Terhal}},\ and\ \bibinfo {author} {\bibfnamefont {C.}~\bibnamefont
  {Vuillot}},\ }\bibfield  {title} {\bibinfo {title} {Roads towards
  fault-tolerant universal quantum computation},\ }\href
  {https://doi.org/10.1038/nature23460} {\bibfield  {journal} {\bibinfo
  {journal} {Nature}\ }\textbf {\bibinfo {volume} {549}},\ \bibinfo {pages}
  {172} (\bibinfo {year} {2017})}\BibitemShut {NoStop}%
\bibitem [{\citenamefont {M\"ott\"onen}\ \emph {et~al.}(2005)\citenamefont
  {M\"ott\"onen}, \citenamefont {Vartiainen}, \citenamefont {Bergholm},\ and\
  \citenamefont {Salomaa}}]{mottonen2005transformation}%
  \BibitemOpen
  \bibfield  {author} {\bibinfo {author} {\bibfnamefont {M.}~\bibnamefont
  {M\"ott\"onen}}, \bibinfo {author} {\bibfnamefont {J.~J.}\ \bibnamefont
  {Vartiainen}}, \bibinfo {author} {\bibfnamefont {V.}~\bibnamefont
  {Bergholm}},\ and\ \bibinfo {author} {\bibfnamefont {M.~M.}\ \bibnamefont
  {Salomaa}},\ }\bibfield  {title} {\bibinfo {title} {{Transformation of
  quantum states using uniformly controlled rotations}},\ }\href
  {https://doi.org/10.5555/2011670.2011675} {\bibfield  {journal} {\bibinfo
  {journal} {Quantum Inf. Comput.}\ }\textbf {\bibinfo {volume} {5}},\ \bibinfo
  {pages} {467} (\bibinfo {year} {2005})}\BibitemShut {NoStop}%
\bibitem [{\citenamefont {Araujo}\ \emph {et~al.}(2021)\citenamefont {Araujo},
  \citenamefont {Park}, \citenamefont {Petruccione},\ and\ \citenamefont
  {Silva}}]{araujo2021divide}%
  \BibitemOpen
  \bibfield  {author} {\bibinfo {author} {\bibfnamefont {I.~F.}\ \bibnamefont
  {Araujo}}, \bibinfo {author} {\bibfnamefont {D.~K.}\ \bibnamefont {Park}},
  \bibinfo {author} {\bibfnamefont {F.}~\bibnamefont {Petruccione}},\ and\
  \bibinfo {author} {\bibfnamefont {A.~J.~d.}\ \bibnamefont {Silva}},\
  }\bibfield  {title} {\bibinfo {title} {{A divide-and-conquer algorithm for
  quantum state preparation}},\ }\href
  {https://doi.org/10.1038/s41598-021-85474-1} {\bibfield  {journal} {\bibinfo
  {journal} {Sci. Rep.}\ }\textbf {\bibinfo {volume} {11}},\ \bibinfo {pages}
  {6329} (\bibinfo {year} {2021})}\BibitemShut {NoStop}%
\bibitem [{\citenamefont {Sun}(2015)}]{sun2015libcint}%
  \BibitemOpen
  \bibfield  {author} {\bibinfo {author} {\bibfnamefont {Q.}~\bibnamefont
  {Sun}},\ }\bibfield  {title} {\bibinfo {title} {Libcint: {An} efficient
  general integral library for gaussian basis functions},\ }\href
  {https://doi.org/10.1002/jcc.23981} {\bibfield  {journal} {\bibinfo
  {journal} {Journal of computational chemistry}\ }\textbf {\bibinfo {volume}
  {36}},\ \bibinfo {pages} {1664} (\bibinfo {year} {2015})}\BibitemShut
  {NoStop}%
\bibitem [{\citenamefont {Sun}\ \emph {et~al.}(2018)\citenamefont {Sun},
  \citenamefont {Berkelbach}, \citenamefont {Blunt}, \citenamefont {Booth},
  \citenamefont {Guo}, \citenamefont {Li}, \citenamefont {Liu}, \citenamefont
  {McClain}, \citenamefont {Sayfutyarova}, \citenamefont {Sharma} \emph
  {et~al.}}]{sun2018pyscf}%
  \BibitemOpen
  \bibfield  {author} {\bibinfo {author} {\bibfnamefont {Q.}~\bibnamefont
  {Sun}}, \bibinfo {author} {\bibfnamefont {T.~C.}\ \bibnamefont {Berkelbach}},
  \bibinfo {author} {\bibfnamefont {N.~S.}\ \bibnamefont {Blunt}}, \bibinfo
  {author} {\bibfnamefont {G.~H.}\ \bibnamefont {Booth}}, \bibinfo {author}
  {\bibfnamefont {S.}~\bibnamefont {Guo}}, \bibinfo {author} {\bibfnamefont
  {Z.}~\bibnamefont {Li}}, \bibinfo {author} {\bibfnamefont {J.}~\bibnamefont
  {Liu}}, \bibinfo {author} {\bibfnamefont {J.~D.}\ \bibnamefont {McClain}},
  \bibinfo {author} {\bibfnamefont {E.~R.}\ \bibnamefont {Sayfutyarova}},
  \bibinfo {author} {\bibfnamefont {S.}~\bibnamefont {Sharma}}, \emph
  {et~al.},\ }\bibfield  {title} {\bibinfo {title} {Pyscf: the python-based
  simulations of chemistry framework},\ }\href
  {https://doi.org/10.1002/wcms.1340} {\bibfield  {journal} {\bibinfo
  {journal} {Wiley Interdisciplinary Reviews: Computational Molecular Science}\
  }\textbf {\bibinfo {volume} {8}},\ \bibinfo {pages} {e1340} (\bibinfo {year}
  {2018})}\BibitemShut {NoStop}%
\bibitem [{\citenamefont {Sun}\ \emph {et~al.}(2020)\citenamefont {Sun},
  \citenamefont {Zhang}, \citenamefont {Banerjee}, \citenamefont {Bao},
  \citenamefont {Barbry}, \citenamefont {Blunt}, \citenamefont {Bogdanov},
  \citenamefont {Booth}, \citenamefont {Chen}, \citenamefont {Cui} \emph
  {et~al.}}]{sun2020recent}%
  \BibitemOpen
  \bibfield  {author} {\bibinfo {author} {\bibfnamefont {Q.}~\bibnamefont
  {Sun}}, \bibinfo {author} {\bibfnamefont {X.}~\bibnamefont {Zhang}}, \bibinfo
  {author} {\bibfnamefont {S.}~\bibnamefont {Banerjee}}, \bibinfo {author}
  {\bibfnamefont {P.}~\bibnamefont {Bao}}, \bibinfo {author} {\bibfnamefont
  {M.}~\bibnamefont {Barbry}}, \bibinfo {author} {\bibfnamefont {N.~S.}\
  \bibnamefont {Blunt}}, \bibinfo {author} {\bibfnamefont {N.~A.}\ \bibnamefont
  {Bogdanov}}, \bibinfo {author} {\bibfnamefont {G.~H.}\ \bibnamefont {Booth}},
  \bibinfo {author} {\bibfnamefont {J.}~\bibnamefont {Chen}}, \bibinfo {author}
  {\bibfnamefont {Z.-H.}\ \bibnamefont {Cui}}, \emph {et~al.},\ }\bibfield
  {title} {\bibinfo {title} {Recent developments in the pyscf program
  package},\ }\bibfield  {journal} {\bibinfo  {journal} {The Journal of
  chemical physics}\ }\textbf {\bibinfo {volume} {153}},\ \href
  {https://doi.org/10.1063/5.0006074} {10.1063/5.0006074} (\bibinfo {year}
  {2020})\BibitemShut {NoStop}%
\bibitem [{\citenamefont {McClean}\ \emph {et~al.}(2020)\citenamefont
  {McClean}, \citenamefont {Rubin}, \citenamefont {Sung}, \citenamefont
  {Kivlichan}, \citenamefont {Bonet-Monroig}, \citenamefont {Cao},
  \citenamefont {Dai}, \citenamefont {Fried}, \citenamefont {Gidney},
  \citenamefont {Gimby} \emph {et~al.}}]{mcclean2020openfermion}%
  \BibitemOpen
  \bibfield  {author} {\bibinfo {author} {\bibfnamefont {J.~R.}\ \bibnamefont
  {McClean}}, \bibinfo {author} {\bibfnamefont {N.~C.}\ \bibnamefont {Rubin}},
  \bibinfo {author} {\bibfnamefont {K.~J.}\ \bibnamefont {Sung}}, \bibinfo
  {author} {\bibfnamefont {I.~D.}\ \bibnamefont {Kivlichan}}, \bibinfo {author}
  {\bibfnamefont {X.}~\bibnamefont {Bonet-Monroig}}, \bibinfo {author}
  {\bibfnamefont {Y.}~\bibnamefont {Cao}}, \bibinfo {author} {\bibfnamefont
  {C.}~\bibnamefont {Dai}}, \bibinfo {author} {\bibfnamefont {E.~S.}\
  \bibnamefont {Fried}}, \bibinfo {author} {\bibfnamefont {C.}~\bibnamefont
  {Gidney}}, \bibinfo {author} {\bibfnamefont {B.}~\bibnamefont {Gimby}}, \emph
  {et~al.},\ }\bibfield  {title} {\bibinfo {title} {Openfermion: the electronic
  structure package for quantum computers},\ }\href
  {https://doi.org/10.1088/2058-9565/ab8ebc} {\bibfield  {journal} {\bibinfo
  {journal} {Quantum Science and Technology}\ }\textbf {\bibinfo {volume}
  {5}},\ \bibinfo {pages} {034014} (\bibinfo {year} {2020})}\BibitemShut
  {NoStop}%
\bibitem [{\citenamefont {Sivarajah}\ \emph {et~al.}(2020)\citenamefont
  {Sivarajah}, \citenamefont {Dilkes}, \citenamefont {Cowtan}, \citenamefont
  {Simmons}, \citenamefont {Edgington},\ and\ \citenamefont
  {Duncan}}]{sivarajah2020t}%
  \BibitemOpen
  \bibfield  {author} {\bibinfo {author} {\bibfnamefont {S.}~\bibnamefont
  {Sivarajah}}, \bibinfo {author} {\bibfnamefont {S.}~\bibnamefont {Dilkes}},
  \bibinfo {author} {\bibfnamefont {A.}~\bibnamefont {Cowtan}}, \bibinfo
  {author} {\bibfnamefont {W.}~\bibnamefont {Simmons}}, \bibinfo {author}
  {\bibfnamefont {A.}~\bibnamefont {Edgington}},\ and\ \bibinfo {author}
  {\bibfnamefont {R.}~\bibnamefont {Duncan}},\ }\bibfield  {title} {\bibinfo
  {title} {t$|$ket$\rangle$: a retargetable compiler for nisq devices},\ }\href
  {https://doi.org/10.1088/2058-9565/ab8e92} {\bibfield  {journal} {\bibinfo
  {journal} {Quantum Science and Technology}\ }\textbf {\bibinfo {volume}
  {6}},\ \bibinfo {pages} {014003} (\bibinfo {year} {2020})}\BibitemShut
  {NoStop}%
\bibitem [{\citenamefont {Nielsen}\ \emph {et~al.}(2020)\citenamefont
  {Nielsen}, \citenamefont {Rudinger}, \citenamefont {Proctor}, \citenamefont
  {Russo}, \citenamefont {Young},\ and\ \citenamefont
  {Blume-Kohout}}]{nielsen2020probing}%
  \BibitemOpen
  \bibfield  {author} {\bibinfo {author} {\bibfnamefont {E.}~\bibnamefont
  {Nielsen}}, \bibinfo {author} {\bibfnamefont {K.}~\bibnamefont {Rudinger}},
  \bibinfo {author} {\bibfnamefont {T.}~\bibnamefont {Proctor}}, \bibinfo
  {author} {\bibfnamefont {A.}~\bibnamefont {Russo}}, \bibinfo {author}
  {\bibfnamefont {K.}~\bibnamefont {Young}},\ and\ \bibinfo {author}
  {\bibfnamefont {R.}~\bibnamefont {Blume-Kohout}},\ }\bibfield  {title}
  {\bibinfo {title} {Probing quantum processor performance with pygsti},\
  }\href {https://doi.org/10.1088/2058-9565/ab8aa4} {\bibfield  {journal}
  {\bibinfo  {journal} {Quantum science and Technology}\ }\textbf {\bibinfo
  {volume} {5}},\ \bibinfo {pages} {044002} (\bibinfo {year}
  {2020})}\BibitemShut {NoStop}%
\bibitem [{\citenamefont {Cross}\ \emph {et~al.}(2022)\citenamefont {Cross},
  \citenamefont {Javadi-Abhari}, \citenamefont {Alexander}, \citenamefont
  {De~Beaudrap}, \citenamefont {Bishop}, \citenamefont {Heidel}, \citenamefont
  {Ryan}, \citenamefont {Sivarajah}, \citenamefont {Smolin}, \citenamefont
  {Gambetta} \emph {et~al.}}]{cross2022openqasm}%
  \BibitemOpen
  \bibfield  {author} {\bibinfo {author} {\bibfnamefont {A.}~\bibnamefont
  {Cross}}, \bibinfo {author} {\bibfnamefont {A.}~\bibnamefont
  {Javadi-Abhari}}, \bibinfo {author} {\bibfnamefont {T.}~\bibnamefont
  {Alexander}}, \bibinfo {author} {\bibfnamefont {N.}~\bibnamefont
  {De~Beaudrap}}, \bibinfo {author} {\bibfnamefont {L.~S.}\ \bibnamefont
  {Bishop}}, \bibinfo {author} {\bibfnamefont {S.}~\bibnamefont {Heidel}},
  \bibinfo {author} {\bibfnamefont {C.~A.}\ \bibnamefont {Ryan}}, \bibinfo
  {author} {\bibfnamefont {P.}~\bibnamefont {Sivarajah}}, \bibinfo {author}
  {\bibfnamefont {J.}~\bibnamefont {Smolin}}, \bibinfo {author} {\bibfnamefont
  {J.~M.}\ \bibnamefont {Gambetta}}, \emph {et~al.},\ }\bibfield  {title}
  {\bibinfo {title} {Openqasm 3: A broader and deeper quantum assembly
  language},\ }\href {https://doi.org/10.1145/3505636} {\bibfield  {journal}
  {\bibinfo  {journal} {ACM Transactions on Quantum Computing}\ }\textbf
  {\bibinfo {volume} {3}},\ \bibinfo {pages} {1} (\bibinfo {year}
  {2022})}\BibitemShut {NoStop}%
\bibitem [{\citenamefont {Javadi-Abhari}\ \emph {et~al.}(2024)\citenamefont
  {Javadi-Abhari}, \citenamefont {Treinish}, \citenamefont {Krsulich},
  \citenamefont {Wood}, \citenamefont {Lishman}, \citenamefont {Gacon},
  \citenamefont {Martiel}, \citenamefont {Nation}, \citenamefont {Bishop},
  \citenamefont {Cross} \emph {et~al.}}]{javadi2024quantum}%
  \BibitemOpen
  \bibfield  {author} {\bibinfo {author} {\bibfnamefont {A.}~\bibnamefont
  {Javadi-Abhari}}, \bibinfo {author} {\bibfnamefont {M.}~\bibnamefont
  {Treinish}}, \bibinfo {author} {\bibfnamefont {K.}~\bibnamefont {Krsulich}},
  \bibinfo {author} {\bibfnamefont {C.~J.}\ \bibnamefont {Wood}}, \bibinfo
  {author} {\bibfnamefont {J.}~\bibnamefont {Lishman}}, \bibinfo {author}
  {\bibfnamefont {J.}~\bibnamefont {Gacon}}, \bibinfo {author} {\bibfnamefont
  {S.}~\bibnamefont {Martiel}}, \bibinfo {author} {\bibfnamefont {P.~D.}\
  \bibnamefont {Nation}}, \bibinfo {author} {\bibfnamefont {L.~S.}\
  \bibnamefont {Bishop}}, \bibinfo {author} {\bibfnamefont {A.~W.}\
  \bibnamefont {Cross}}, \emph {et~al.},\ }\bibfield  {title} {\bibinfo {title}
  {Quantum computing with qiskit},\ }\bibfield  {journal} {\bibinfo  {journal}
  {arXiv preprint arXiv:2405.08810}\ }\href
  {https://doi.org/10.48550/arXiv.2405.08810} {10.48550/arXiv.2405.08810}
  (\bibinfo {year} {2024})\BibitemShut {NoStop}%
\end{thebibliography}%

\appendix

\section{Circuit compilation}\label{app:circuits}
SVB takes, as its input, a circuit $c$ that has been \emph{fully compiled} for a particular quantum computer $\qc$. In this appendix we provide additional details on what this means, and discuss it in the context of a quantum algorithm or application (which is the primary setting in which we anticipate SVB will be employed). A circuit $c$ is fully compiled for $\qc$ if each qubit in $c$ is assigned to a physical qubit in $\qc$ and if $c$ contains only the native gates of $\qc$, including only interacting qubits that are coupled in $\qc$'s connectivity graph. There is some ambiguity about precisely what a quantum computer's ``native gates'' are (e.g., arbitrary single qubit gates are implementable by two rotations by $\pi/2$ around the $X$ axis together with three rotations around the $Z$ axis, which are typically implemented virtually, so an arbitrary single qubit gate might be considered a native operation). These somewhat arbitrary choices impact the exact meaning of a fully compiled circuit, and then impact the depth of a fully compiled circuit. This, however, has no important consequences for SVB---as the depth of a circuit snippet from $c$ will typically be interpreted relatively to $c$'s depth---and so we do not need to be pedantic about precisely how the native gate set is defined.

The reason why we require that $c$ is fully compiled is that compiling of the circuit snippets SVB samples from $c$ is not permitted (except for trivial compilations, like replacing an $X$ gate with two $X_{\pi/2}$ gates). Unless $c$ is already fully compiled, this would prevent the execution of those snippets. The precise constraint we set on the compilations of our snippets can be formalized using the concept of \emph{compilation barriers}: each layer of $c$ is separated by a compilation barrier, and those barriers persist in our snippets. We demand no compilation of our snippets because otherwise it is possible that compilations could reduce the depth or change other characteristics of those snippets (e.g., adding ancillary qubits), making the fidelities of those snippets less likely to be predictive of the fidelity of $\qc$. 

To create a fully compiled $c$ for SVB there is typically a multi-step process, which is illustrated in the top row of Fig.~\ref{fig:SVB-overview}. SVB is, we anticipate, most interesting for the case of quantifying progress towards solving some computational problem with a quantum computer. Therefore, we must first select a problem instance and a quantum algorithm for solving this problem. Then all tunable parameters of the algorithm must be chosen (e.g., the number of Trotter steps in an algorithm that uses Trotterization). These steps then typically produce one or more ``high-level'' circuits: circuits that are expressed in terms of subroutines that cannot be directly implemented on most or all quantum computing hardware (e.g., the quantum Fourier transform, or $n$-qubit Toffoli gates). This circuit then needs to be compiled to be executable on the particular quantum computer $\qc$ in question, which is a step that is often broken down into multiple stages, such as qubit mapping/routing. Note, however, that SVB can be separately applied to circuits created with different choices at one or more stages (e.g., different algorithm choices for the same problem) and, by doing so, the performance with different choices can be compared.

\section{Hamiltonian simulation SVBs}\label{app:demonstration-details}
In this appendix we provide additional details of the Hamiltonian simulation SVBs presented in the main text.

\subsection{Hamiltonian block encoding circuits}\label{app:block-encoded-circs}
We briefly review the block encoding framework \cite{low2017hamiltonian, chakraborty2018power, gilyen2019quantum} and, in particular, the LCU approach to this block encoding \cite{childs2012hamiltonian}, which is what we use in our demonstration of SVB.

\begin{figure}[t]
    \centering
    \begin{quantikz}
    \lstick{\ket{l}} & \qw\qwbundle{m} &\gate{U_{\rm PREP}} & \gate[wires=2]{U_{\rm SEL}} & \gate{U_{\rm UNPREP}^\dagger} & \qw & \\
    \lstick{\ket{\Psi}} & \qw\qwbundle{n} & \qw &  \qw & \qw & \qw &
    \end{quantikz}
    \caption{General form of an LCU circuit.}
    \label{fig:LCU_circuit}
\end{figure}
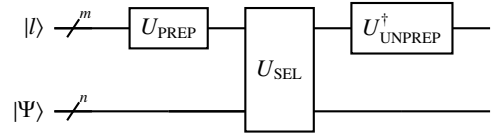

In general, a unitary $U_A$ is an $(\alpha, m, \epsilon)$-block encoding of $A$ if
\begin{align}
    \left\Vert A - \alpha \left(\langle 0\vert^{\otimes m} \otimes I_n\right) U_A
        \left(\vert0\rangle^{\otimes m} \otimes I_n\right) \right\Vert \leq \epsilon,
\end{align}
where $\alpha$ serves to scale $A$ such that the norm of $U_A = 1$, $n$ and $m$ are the sizes of the system and auxiliary registers, respectively, and $\epsilon$ bounds the error of the block encoding. Suppose that the operator $A$ can be expanded as a linear combination of $L$ unitaries,
\begin{equation}
    A = \sum_{l=0}^{L} c_l U_l,
        \label{eqn:LCU_A}
\end{equation}
where $L$ necessarily sets the size of auxiliary register such that $L < 2^m - 1$.
We can define a pair of state preparation unitaries $U_{\rm PREP}$ and $U_{\rm UNPREP}$ and a select unitary $U_{\rm SEL}$ as
\begin{align}
U_{\rm PREP}\vert0\rangle^{\otimes m} &= \sum_{l=0}^{L} a_l\vert l\rangle
    \label{eqn:U_PREP},\\
U_{\rm UNPREP}\vert0\rangle^{\otimes m} &= \sum_{l=0}^{L} b_l\vert l\rangle
    \label{eqn:U_UNPREP},\\
U_{\rm SEL} &= \sum_{l=0}^{L} \vert l\rangle\langle l\vert \otimes U_l
    \label{eqn:U_SEL},
\end{align}
such that
\begin{equation}
\sum_{l=0}^{L} \left\vert a_l^* b_l - \frac{c_l}{\Vert c\Vert_1}\right\vert < \epsilon_1.
\label{eqn:spp_condition}
\end{equation}
The LCU encoding given by
\begin{align}
    U_A &= (U_{\rm UNPREP}^\dagger\otimes I_n) U_{\rm SEL} (U_{\rm PREP}\otimes I_n)
        \label{eqn:LCU_U}
\end{align}
serves as a $(\Vert c \Vert_1, m, \epsilon_1)$-block encoding of $A$. The circuit for this LCU encoding is depicted in Fig. \ref{fig:LCU_circuit}.

In the case where all $c_l \geq 0$, the state preparation pair can be reduced to a single unitary with real coefficients
\begin{align}
a_l = b_l = \sqrt{\frac{c_l}{\Vert c\Vert_1}}.
    \label{eqn:U_PREP_coeffs_positive}
\end{align}
Alternatively, in the case where some $c_l < 0$, the state preparation pair unitaries must have coefficients $a_l$ and $b_l$ such that their inner product yields the requisite negative sign. One simple option is to set $a_l = -b_l$, which corresponds to flipping the sign of certain rotation angles in $U_{\rm UNPREP}$.

\begin{table}[tbp]
    \centering
    \begin{tabular*}{0.4\textwidth}{c|@{\extracolsep{\fill}}ccc}
        \toprule
        Model  & System  & Number of & Auxiliary \\
        System & Qubits & LCU Terms & Qubits \\
        & ($n$) & ($L$) & ($2m-1$) \\
        \colrule
        H$_2$ & 4 & 15 & 7  \\
        HeH$^+$ & 4 & 27 & 9\\
        LiH & 8 & 105 & 13 \\
        \botrule
    \end{tabular*}
    \caption{Number of qubits and terms needed for the LCU encoding of each model system. Auxiliary qubit counts include those needed for unary iteration.}
    \label{tab:mol-widths}
\end{table}

\begin{figure}[pt]
 \centering
    \begin{quantikz}[column sep=0.1cm]
    \lstick{\ket{l_2}} & \gate{Y_{\theta^0}} & \octrl{1} & \ctrl{1} & \octrl{1} & \octrl{1} & \ctrl{1} & \ctrl{1} & \qw \\
    \lstick{\ket{l_1}} & \qw &  \gate{Y_{\theta_{0}^1}} & \gate{Y_{\theta_{1}^1}} & \octrl{1} & \ctrl{1} & \octrl{1} & \ctrl{1} &  \qw \\
    \lstick{\ket{l_0}} & \qw &  \qw &  \qw &  \gate{Y_{\theta_{0}^2}} & \gate{Y_{\theta_{1}^2}} & \gate{Y_{\theta_{2}^2}} & \gate{Y_{\theta_{3}^2}} & \qw
    \end{quantikz}
\newline
\vspace{0.5cm}
\newline
    \begin{quantikz}[column sep=0.2cm]
    \lstick{\ket{l_2}}
      &\qw &\qw&\octrl{1} & \octrl{1}& \octrl{1}& \octrl{1} & \ctrl{1}& \ctrl{1}& \ctrl{1}& \ctrl{1} &  \qw\\
    \lstick{\ket{l_1}}
     & \qw &\qw& \octrl{1}& \octrl{1} & \ctrl{1} & \ctrl{1}& \octrl{1}& \octrl{1}& \ctrl{1}& \ctrl{1}& \qw\\
    \lstick{\ket{l_0}}
     & \qw &\qw& \octrl{1} & \ctrl{1}& \octrl{1} & \ctrl{1}& \octrl{1}& \ctrl{1} & \octrl{1}& \ctrl{1}& \qw\\
    \lstick{\ket{\Psi}}
      &\qwbundle{n} &\qw& \gate{U_0} & \gate{U_1} & \gate{U_2} & \gate{U_3}& \gate{U_4}& \gate{U_5}& \gate{U_6} & \gate{U_7}& \qw
    \end{quantikz}
    \caption{\textbf{Prepare (top) and select (bottom) circuits}. An uncompiled LCU circuit consists of a prepare circuit and a select circuit. Here we show these circuit for an 8-term LCU circuit. The uncompiled prepare circuit (top) consists of controlled rotations with $Y_{\theta_i^j} \equiv R_Y(\theta_i^j)$ where $i$ and $j$ denote the controlling state and number of preceding qubits, respectively. The select unitary (bottom) consists of multi-controlled unitaries conditioned on the state of the auxiliary register.}
    \label{fig:multicontrolled_Uprep_Usel}
\end{figure}
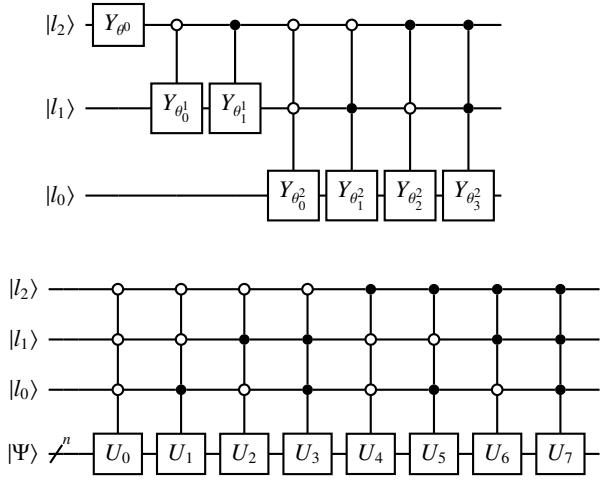

An LCU circuit can be expressed in terms of many multi-controlled gates, as shown in the example of Fig.~\ref{fig:multicontrolled_Uprep_Usel}.  The prepare circuit can be improved by using an efficient decomposition of uniformly controlled rotations that only consist of $R_Y$ (a rotation around the $Y$ axis) and CNOT gates \cite{mottonen2005transformation}. The select circuit can be improved through a procedure known as ``unary iteration" where additional qubits are introduced to store partial computations of the state in the auxiliary register \cite{babbush2018encoding}. For an LCU with $m$ auxiliary qubits, unary iteration introduces $m-1$ additional qubits but results in a massive reduction in depth of the final compiled circuit, relative to a naive implementation like the one in Fig.~\ref{fig:multicontrolled_Uprep_Usel}. Further implementation details for the prepare oracle are described below.

\subsubsection{Angles for State Preparation}\label{app:state_prep_angles}
State preparation pair unitaries using uniformly controlled rotations such as the one shown in Fig.~\ref{fig:multicontrolled_Uprep_Usel} are defined by their $R_Y$ rotation angles $\theta_i^j$, where $i$ and $j$ denote the controlling state and number of preceding qubits, respectively. Each angle can be computed \cite{mottonen2005transformation} as
\begin{equation}
    \theta_{i}^j = 2\arcsin\left(\frac{\sqrt{\sum_{k=0}^{2^j-1} \left|a_{(2i+1)2^{j} + k}\right|^2 }}
		{\sqrt{\sum_{k=0}^{2^{j+1}-1} \left|a_{i2^{j+1} + l}\right|^2 }}\right),
	\label{eqn:theta_mottonen}
\end{equation}
where $a_i$ are the coefficients of the prepared state from Eq.~\eqref{eqn:U_PREP}. For $m$ auxiliary qubits, the indexing quantities can vary as $j=0,1,\cdots,m-1$ and $i=0,1,\cdots,2^{m-j-1}$.

However, one can also compute the angles given in Eq.~\eqref{eqn:theta_mottonen} using a binary trie where each node corresponds to a bit string and left/right children append a 0/1 to the bit string, respectively. Recent work on parallel implementations of state preparation unitaries \cite{araujo2021divide} has utilized similar data structures, and we use their terminology of a ``state tree" to denote a binary trie that stores coefficient information. Leaves hold the final coefficients, internal nodes hold partial norms of their children needed to calculate the $R_Y$ angles, and the root node necessarily holds a norm of 1. Each node's location provides a bit string of the necessary control values for the multi-controlled-$R_Y$ gates.

The rotation angles $\theta_i^j$ can now be calculated by navigating the state tree according to the bit string representation of $i$, which should reside on the $j$\textsuperscript{th} layer of the state tree. The current node's partial norm will take the place of the denominator in Eq.~\eqref{eqn:theta_mottonen} while the right child's partial norm takes the place of the numerator. Calculating all rotation angles needed for a full state preparation using the state tree is more efficient than using Eq.~\eqref{eqn:theta_mottonen} because there will be fewer partial norm summations.

The state tree can also be used to easily compute which angles will need to be flipped between $U_{\rm PREP}$ and $U_{\rm UNPREP}$ to satisfy Eq.~\eqref{eqn:spp_condition}. The needed sign can be constructed in a bottom-up approach using the following rules: 1) each leaf takes the sign of its coefficient, 2) a node's angle must be flipped in $U_{\rm UNPREP}$ if the signs of its children don't match, and 3) a node inherits the sign of its right child.

\subsection{Implementation details}\label{app:implementation}
The LCU coefficients were computed with PySCF \cite{sun2015libcint,sun2018pyscf,sun2020recent}. OpenFermion \cite{mcclean2020openfermion} was used to perform the Jordan-Wigner and Bravyi-Kitaev mappings, utilizing OpenFermion-PySCF to interface between the two packages. The LCU circuits were then expressed in PyTket \cite{sivarajah2020t} using high-level operations such as multi-controlled Pauli strings before being converted to the TK1 and CNOT gate set. PyTket's full peephole optimization pass was applied to the circuit before the noise-aware mapping and routing passes were used to embed the qubits and enforce the correct CNOT connectivity for the target IBM Q device. The circuit was rebased once more into the U3 and CNOT gate set, which only requires a minor adjustment to the angles in the one-qubit gates, before being passed to PyGSTi \cite{nielsen2020probing}. Subcircuits of each target width and depth were sampled from the full circuit before the circuit mirroring procedure was carried out on each subcircuit. The U3 gates were decomposed into $\pi/2$ rotations around the $X$ axis (IBM's``sx'' gate) and $R_Z$ gates, converted into QASM \cite{cross2022openqasm}, and submitted to IBM Q using Qiskit \cite{javadi2024quantum}.
\end{document}